\documentstyle[aps,epsf]{revtex} 
 
\begin{document} 
\draft 

\title{Unstable decay and state selection} 
 
\author{Alan McKane$^1$ and Martin Tarlie$^2$} 
 
\address{ 
$^1$Department of Theoretical Physics, University of Manchester, 
Manchester M13 9PL, England 
} 

\address{ 
$^2$James Franck Institute, University of Chicago, 5640 South Ellis Avenue,
Chicago IL 60637 
} 

\maketitle   
 
\begin{abstract}
The decay of unstable states when several metastable states are available 
for occupation is investigated using path-integral techniques. Specifically,
a method is described which allows the probabilities with which the metastable
states are occupied to be calculated by finding optimal paths, and 
fluctuations about them, in the weak noise limit. The method is illustrated 
on a system described by two coupled Langevin equations, which are found in 
the study of instabilities in fluid dynamics and superconductivity. The 
problem involves a subtle interplay between non-linearities and noise, and a 
naive approximation scheme which does not take this into account is shown to 
be unsatisfactory. The use of optimal paths is briefly reviewed and then 
applied to finding the conditional probability of ending up in one of the 
metastable states, having begun in the unstable state. There are several 
aspects of the calculation which distinguish it from most others involving 
optimal paths: (i) the paths do not begin and end on an attractor, and 
moreover, the final point is to a large extent arbitrary, (ii) the interplay 
between the fluctuations and the leading order contribution are at the heart 
of the method, and (iii) the final result involves quantities which are not 
exponentially small in the noise strength. This final result, which gives 
the probability of a particular state being selected in terms of the 
parameters of the dynamics, is remarkably simple and agrees well with the 
results of numerical simulations. The method should be applicable to similar
problems in a number of other areas such as state selection in lasers, 
activationless chemical reactions and population dynamics in fluctuating
environments.
\end{abstract} 
 
\vspace{0.3in} 
 
\section{Introduction} 
The decay of metastable states, due to thermal or other random fluctuations, 
is a phenomenon seen in many diverse areas of science, and consequently has
a huge literature associated with it. In the simplest cases, where a potential
can be defined and the states assigned a particular value of this potential, 
the decay process can be viewed as noise activation over the potential 
barriers that separate the metastable state under consideration from all of 
the other accessible metastable states of the system. The average time taken 
to escape from a potential well (i.e. for the state to decay) is of the 
order of $\exp{\Delta V/D}$, where $\Delta V$ is the height of the barrier 
to be surmounted and $D$ is the strength of the noise. Thus the picture we 
have in this case, is of a set of metastable states, with transitions 
between them which occur with probabilities which depend on the nature of 
the potential between the states and on the strength of the noise.     

By contrast, the decay of unstable states, although a similarly widespread 
phenomenon, has been studied much less. In terms of the above picture, the
system starts at or near a maximum of the potential and makes transitions to 
the accessible metastable states of the system with various probabilities,
which as before depend on the nature of the potential between the unstable 
and metastable states and on the noise. In an earlier paper\cite{I} we 
introduced a scheme which enabled us to calculate the probabilities with 
which the various metastable states are selected. Our aim here is to extend
this work, by giving a fuller presentation of the ideas and techniques 
involved, justifying some of the earlier approximations which were made and
discussing the link with real systems in more detail.

The phenomenon of the selection of metastable states, from states which have
become unstable, is at the heart of pattern formation and the origin of
complex structures, with the selection of non-trivial states being governed 
partly by the deterministic dynamics and partly by the noise acting on the 
system. Thus the picture painted above, while a simplified version of this
general scenario, contains many of its essential features. In fact, the
above structure can be derived from the equations describing the entire 
system by focusing on the unstable modes and the modes which have the 
potential to be selected, and treating all of the other modes as background
noise. The resulting dynamics then comprises a small number of coupled 
ordinary differential equations acted on by noise. If this flow is potential,
then the above picture is recovered; if not, our techniques are still 
applicable, but it becomes more difficult to visualize.

State selection of the kind we have been describing is ubiquitous. In fluid
dynamics, it appears when Rayleigh-B\'{e}nard convection rolls of given
wavenumber are formed after the decay of unstable ones. A set of equations
describing the non-linear coupling between parallel rolls of different 
wave numbers may be derived\cite{fd_1}, and when the effects of the other
modes are incorporated\cite{fd_2}, a set of coupled stochastic differential 
equations of the kind mentioned above are generated. In superconductivity,
exactly the same equations as in the above example govern state selection
in, for example, narrow superconducting rings. This is simply because the 
amplitude equation which governs the instabilities in the fluid dynamics 
example is nothing but the Ginzburg-Landau equation for a 
superconductor\cite{super_1}. A mode truncation then gives exactly the 
same set of coupled differential equations acted upon by noise\cite{super_2}.
In chemical kinetics, it has become clear in the last decade or two, that 
there are many important chemical reactions in which a barrier to the 
formation of an excited state is not present. Examples of these 
activationless reactions\cite{crd_1} are the electronic relaxation of 
triphenylmethane dyes and barrierless electron transfer in solution. The 
potential discussed above is a reaction potential energy surface in this 
case\cite{crd_2}. In lasers, such as the ring dye laser\cite{las_1}, decay 
of an unstable mode to metastable ones can occur under the right 
operating conditions. The coordinates in this case are the mode amplitudes 
and Langevin-type equations are derived within the semiclassical theory 
of the laser\cite{las_2}. In population dynamics, the Gause model of
two competing species\cite{pd_1} again falls into this class. 
When the competition is played out with a fluctuating environment, the 
resulting stochastic differential equations once again fall into the 
generic class that we have been discussing\cite{pd_2}. It should, however, 
be noted that in this case there is no potential and moreover the noise 
is multiplicative.

Although many of these phenomena have been known for some time, the 
investigation of the state selection aspect has been hampered by the lack of
a suitable calculational tool. In the case where a potential $V$ exists, the 
noise is additive and only two modes, $x$ and $y$, are considered, the 
equations take the form
\begin{eqnarray}
\dot{x} & = & - \frac{\partial V}{\partial x} + \eta_{x}(t) \nonumber \\
\dot{y} & = & - \frac{\partial V}{\partial y} + \eta_{y}(t)\,,
\label{Langevin}
\end{eqnarray}
where $\eta_x$ and $\eta_{y}$ are white noises of strength $D$. But, even in 
this, the simplest non-trivial case, these equations are difficult to study
mainly because, as we shall discuss later, in the region where state selection
occurs, the coupled nature of the equations, their non-linearity and the 
noise, are all important. However, as we have shown\cite{I}, there is a 
method that can take all these aspects into account, and that is the 
path-integral formulation of stochastic dynamics. This method succeeds where 
others fail because the equations can be represented as a path-integral 
{\it without approximation}; systematic approximation techniques for 
path-integrals developed over the years can then be used as the basis of a 
calculational scheme. 

Most of the previous theoretical work on this problem has been limited to
systems with one degree of freedom, governed by the single equation
$\dot{x} = - V'(x) + \eta(t)$, because many of the complexities mentioned
above are not present in this one-dimensional case. Suzuki and co-workers
developed a theory for the decay of an unstable state in one dimension in a
series of papers\cite{suz}, as did a number of other 
authors\cite{Weiss,ko,dekker,bzd,Hu,hd,pc_1}. However, the one-dimensional 
theory, although much easier to deal with, has none of the subtleties inherent
in state-selection in higher dimensions: the decay is simply either to the 
right or to the left. Some studies\cite{apu,ys,pc_2} purported to go beyond 
one dimension, but in fact considered spherically symmetric potentials, so
that the problem could be reduced to a quasi-one-dimensional problem in
the radial coordinate. Once again, the resultant structure is not rich enough 
to address the question of state selection. Other approaches, such as 
replacing the stochastic differential equations by the corresponding 
deterministic equations, but with random initial conditions\cite{hhg}, are
unsatisfactory in other ways. Probably the investigation nearest to our
own has been by Mangel\cite{man}, however his main interest was not state 
selection.

The plan of the paper is as follows. In Sec. II we discuss the assumptions
underlying the picture of state selection which we have outlined and 
introduce a generic model which we use to describe our calculational
scheme in detail. An important aspect of our method is the realization
that, typically, a state is selected well before the system reaches this
chosen state. This fact is used to simplify the problem in Sec. III, 
where it is also shown that naive calculational prescriptions, based on 
linearization of the initial dynamics, fail. A more systematic approach 
based on path-integral techniques is introduced in Sec. IV and the 
relevant optimal path is determined. This gives the leading order 
contribution in the limit $D \rightarrow 0$. The next to leading order 
contributions are determined in Sec. V. The results of this analytic 
approach are given and compared to Monte Carlo simulations in Sec. VI and our 
conclusions are presented in Sec. VII. There are two appendixes. Appendix 
A describes some of the elementary approaches described in Sec. III
in more detail. Appendix B contains technical aspects relating to the 
determination of the $y$-coordinate of the optimal path and the calculation 
of the action of that path. 

\section{General concepts}

We have already given several examples of situations where state selection,
following the decay of an unstable state, occurs. In this section we will 
explore the different ways in which the initial state of the system could 
arise, that is, the origin of the unstable state, and give an intuitive
description of the decay and subsequent state selection, which will form the 
basis of our analytical approach.

An unstable state may arise by mechanisms which are either non-adiabatic
or adiabatic. In the non-adiabatic case, the system, in an initially 
(meta)stable state, is transported very rapidly to an unstable state. The 
time scale for the transition from the stable to unstable state is much 
more rapid than the characteristic time scale defining the natural dynamics 
of the system. In this context the transition from the stable to the unstable
state can be ignored and we simply characterize the system as having 
been prepared in an unstable state. Examples of this type were mentioned in 
Sec. I and include chemical reactions with no activation barrier and 
population dynamics in a fluctuating environment. Quasi-one dimensional 
superconductors\cite{super_2} are another example. In the case of the
chemical reactions, an ultra-fast light pulse (a femtosecond laser) pumps 
molecules into an excited state. In terms of potential surfaces, the 
molecules are initialized in an unstable state on the reactive surface. 
The subsequent reaction can be viewed as a nuclear rearrangement on 
this reactive potential energy surface. The evolution of the reaction, 
that is the relative preponderance of reactants and products, can be
studied using other ultrafast lasers. In the population dynamics example,
the population is initialized so as to consist of only a few individuals of 
each species. During the initial period the population of both species grow 
exponentially, independently of the other, until competition drives the 
population of one of the species to zero.

In the adiabatic case, an unstable state may arise when a (meta)stable
state is transported through the instability over a time scale that is slow 
compared to the time scale characterizing the natural dynamics of the system.
An example of this process is provided by a quasi one-dimensional 
superconductor as studied in Ref.\cite{te}. In this case, the system is 
driven by a voltage source that accelerates the supercurrent. However, this 
acceleration cannot continue indefinitely and the system becomes unstable, 
a situation associated with the critical current of the superconductor. 
Once unstable, the relaxation process occurs via ``phase slips'' in which 
spatially localized regions of the wire temporarily lose their
superconducting properties and carry ``normal'', i.e. non-superconducting,
current. This process dissipates the excess energy and the system relaxes 
back to a locally stable state. Mesoscopic wires, i.e. wires that 
are not in the thermodynamic limit, have a finite 
number of discrete metastable stationary current carrying states. State
selection in this case is characterized by the competition between these
metastable current carrying states. The superconductivity instability 
described above is an example of the Eckhaus instability. 

Eckhaus instabilities arise in many physical systems in addition to quasi 
one-dimensional superconductors, including fluids, nematic liquid
crystals and lasers. In these systems, stationary one-dimensional periodic 
patterns are stable for a range of wave vectors, $Q$. For instance, in the 
case of the common Eckhaus instability, stationary solutions exist for 
$Q^{2}< 1$, but are only stable if $Q^{2}< 1/3$. By changing the control 
parameter slightly, $Q$-states with $Q^{2}< 1/3$ may be shifted into the 
unstable $Q^{2}> 1/3$ regime. The essential features of the subsequent 
changes which follow from this may be understood by performing a 
mode-decomposition of the relevant amplitude equation and keeping only the 
previously stable mode and the destabilizing modes with the largest growth 
rates\cite{super_2}.An adiabatic elimination of the unstable mode then 
leaves us with coupled ordinary differential equations for the amplitudes 
of the destabilizing modes. This adiabatic assumption is equivalent to 
assuming that the form of these differential equations does not change on 
the time-scale of state selection, i.e. in the time taken for one of the 
destabilizing modes to outcompete the others\cite{te}

An obvious question which now arises is the following: how do we know that 
the system is initialized in a state where the amplitudes of the competing 
modes, final products, etc are zero? Or expressed in the topographical
terminology of Sec. I: How do we know that we begin exactly at the top
of the hill? Here we are assuming that the small non-zero amplitude, required
to initiate the growth, is induced by fluctuations i.e. by the noise. Once
again, there are several possibilities. It may be that in every realization 
of the stochastic process the system starts near, but not necessarily at the 
origin. As long as there is no bias favoring any one of the competing modes,
an average over many realizations of the process will give the same results 
as if the system had started at the origin in each case. Alternatively, the 
initial probability distribution may be so sharply peaked about the origin,
that any small initial bias is irrelevant to the ultimate choice of state
made by the system. Yet another possibility is that it is just not possible 
to start exactly at the origin (the example taken from population dynamics is
a illustration of this). In these cases the sensitivity of the final result 
to changes in the initial starting point will have to be carefully 
investigated. In addition to all of these specific reasons, on general 
grounds beginning at the origin seems very natural, since it simply means 
that none of the final products or final modes are initially present. 

Since there appear to be good reasons to initialize the system at the origin 
in many cases, it will be assumed in the subsequent theoretical development. 
In all of the examples given in Sec. I, the competing modes grow 
exponentially for an initial period, without any significant interaction 
with each other. This corresponds to a linear growth law of the form 
$\dot{x}_{l}=\alpha_{l}x_{l}$, where $x_{l}$ is the amplitude and 
$\alpha_{l}$ the growth rate of the $l$th mode. Clearly, the main interest 
from a state-selection point of view, is in the nature of the non-linear 
interaction terms between the modes. Since the purpose of this paper is to 
give a clear and intuitive introduction to our calculational scheme, we 
shall choose to illustrate our method on an example where only two competing 
modes are present, and where the interaction terms may be derived from a 
potential. In other words, the governing equations will have the form 
(\ref{Langevin}). Obviously, two is the minimum number of modes we need in 
order to discuss state selection, but having only two modes moving on a
potential surface has the advantage that we can describe our method using 
the pictorial language of ``hills and valleys'' which we have already
adopted on several occasions. Furthermore, since as we have already 
mentioned, the equations derived in both the fluid dynamics and 
superconductivity examples are identical\cite{fd_1,super_2} we will work 
with these. They are  
\begin{eqnarray}
\dot{x} & = & \alpha x - \gamma x y^{2} - \delta x^{3} + \eta_{x}(t) 
\label{x_eqn} \\
\dot{y} & = & \beta y - \gamma y x^{2} - \epsilon y^{3} + \eta_{y}(t)\,,
\label{y_eqn}
\end{eqnarray}
where $x$ and $y$ are the two competing modes with growth rates $\alpha$ and
$\beta$ respectively. All five parameters characterizing the model 
($\alpha, \beta, \gamma, \delta$ and $\epsilon$) are assumed to be positive.
The noise terms $\eta_{x}(t)$ and $\eta_{y}(t)$ are taken to be Gaussian 
random variables with zero mean and 
\begin{equation}
\langle \eta_{i}(t) \eta_{j}(t') \rangle = 2D\delta(t - t')\,,
\label{noise}
\end{equation}
where $i$ and $j$ are either $x$ or $y$ and $D$ is the noise strength.  
Equations (\ref{x_eqn}) and (\ref{y_eqn}) fall into the class of those given 
by (\ref{Langevin}), since they are derivable from the potential
\begin{equation}
V(x,y) = - \frac{\alpha}{2}x^{2} - \frac{\beta}{2}y^{2} + 
\frac{\gamma}{2}x^{2} y^{2} + \frac{\delta}{4}x^{4} + 
\frac{\epsilon}{4}y^{4} \,.
\label{potential}
\end{equation}
We wish, however, to stress that the method is not restricted to systems with
only two modes nor is the existence of a potential a prerequisite. This will 
become clear as the method is explored in subsequent sections. 

It is interesting to note that the simple generalization of the model in 
which the strengths of the noises $\eta_{1}(t)$ and $\eta_{2}(t)$ are both
of the same order, but not equal, already leads to a more complicated 
situation. If we suppose that the strengths of these noises are $D_{1}$ and
$D_{2}$ respectively, then we may introduce $\zeta_{i}(t), i=1,2$ such
that the correlation function of these new noise terms is exactly 
(\ref{noise}), by writing $\eta_{i}(t)=\rho_{i}^{1/2} \zeta_{i}(t)$ where 
the $\rho_{i}$ are constants given by $D_{i}=\rho_{i} D$. This rescaling of 
the noise may, in turn, be eliminated from Eqns. (\ref{x_eqn}) and 
(\ref{y_eqn}) by rescaling $x(t)$ and $y(t)$ by defining new variables $x_1$ 
and $x_2$ via $x(t)=\rho_{1}^{1/2} x_{1}(t)$ and 
$y(t)=\rho_{2}^{1/2} x_{2}(t)$. New constants replacing $\delta$ and
$\epsilon$ which absorb this change of scale can easily be defined, but the
interaction terms $\gamma x y^{2}$ and $\gamma y x^{2}$ become 
$\gamma_{1 2}x_{1}x_{2}^{2}$ and $\gamma_{2 1}x_{2}x_{1}^{2}$ respectively,
with $\gamma_{1 2} \neq \gamma_{2 1}$. Equations such as this are not 
derivable from a potential, nevertheless minor modications of our method are 
applicable to this case.

\begin{figure}[tbh]
\begin{center}
\leavevmode
\hbox{\epsfxsize=0.8\columnwidth \epsfbox{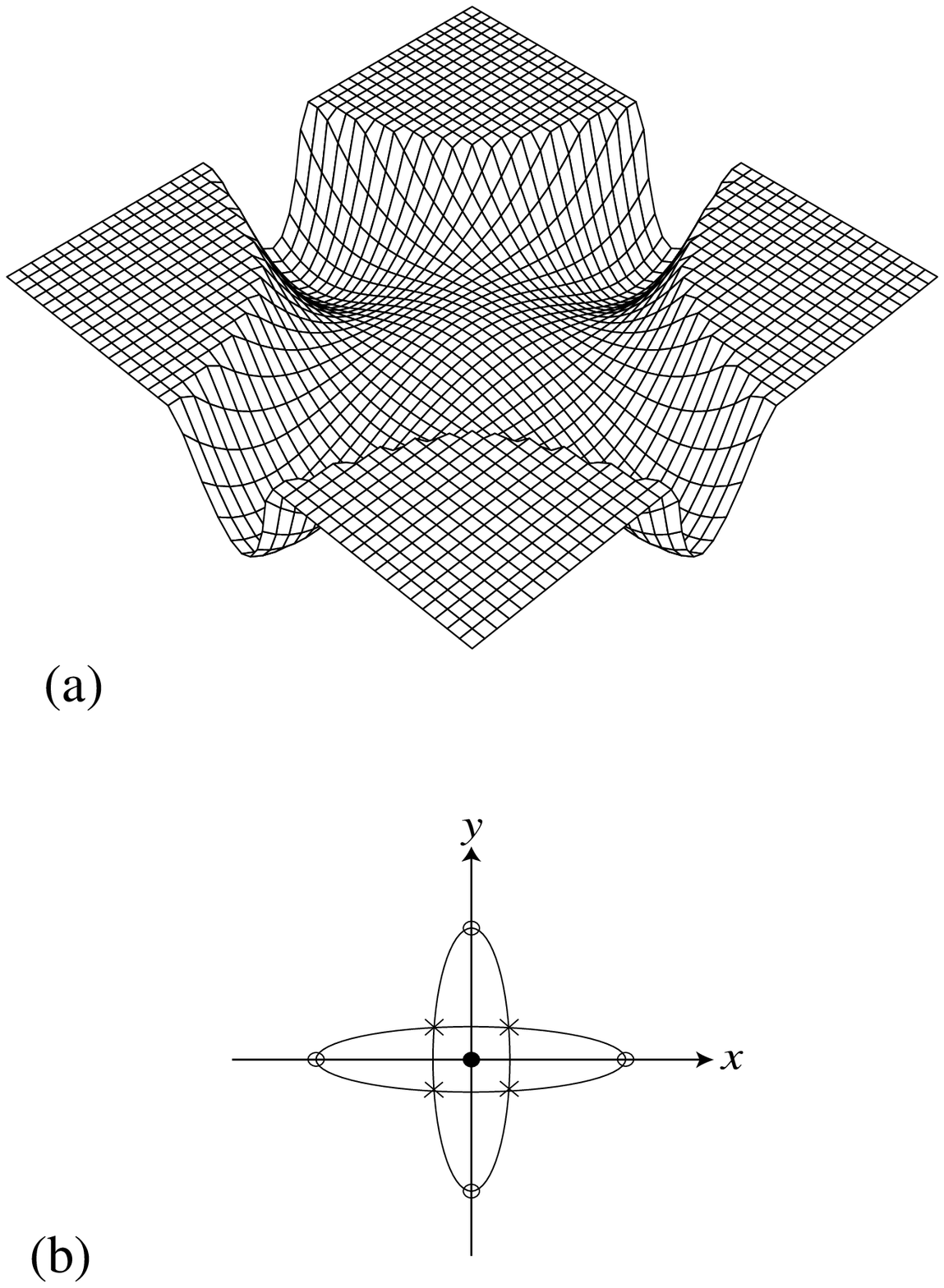}}
\end{center}
\caption{(a) Illustrative plot of $V(x,y)$ with $\alpha = \beta = 2$, 
$\gamma = 4$, $\delta = \epsilon = 1/5$. (b) Contours of zero force.}
\label{Fig1}
\end{figure}

The potential $V(x,y)$ for a particular choice of parameters is shown in
Fig \ref{Fig1}(a). 

Before discussing the dynamics of the system, we have to identify the extrema
of the potential. These are given by the intersection of the two sets of 
curves $\partial V/\partial x = 0$ and $\partial V/\partial y = 0\,$ or, 
explicitly, by the intersection of 
$\{ x=0, \delta x^{2} + \gamma y^{2} = \alpha\}$ and   
$\{ y=0, \gamma x^{2} + \epsilon y^{2} = \beta\}$. These two ellipses, along
with the $x$ and $y$-axes, are shown in Fig \ref{Fig1}(b). By comparing 
Figs \ref{Fig1}(a) and Fig \ref{Fig1}(b) we see that there is (i) a maximum 
at the origin (denoted by a filled circle), (ii) four minima at the 
intersection of the ellipses and the axes (denoted by open circles), and (iii)
four saddle points at the intersection of the two ellipses (denoted by 
crosses). The precise positions of these extrema are:
\begin{eqnarray} 
{\rm Minima}: \ \ (x,y) & = & \left( \pm \left( \frac{\alpha}
{\delta}\right)^{1/2}, \, 0 \right)\ \ ; \ \ (x,y) =  \left( 0, \ \pm 
\left( \frac{\beta}{\epsilon}\right)^{1/2} \right)\,. 
\label{minima} \\ \nonumber \\
{\rm Saddles}: \ \ (x,y) & = & \frac{1}{\sqrt{\gamma^{2} - \delta\epsilon}}
\left( \pm \sqrt{\beta\gamma - \alpha\epsilon}, \ \pm 
\sqrt{\alpha\gamma - \beta\delta} \right)\,. 
\label{saddles}
\end{eqnarray}  
In fact, there are some conditions which need to be imposed on the parameters
of the model if saddle points are to exist. These are that 
$\alpha\gamma - \beta\delta$, $\beta\gamma - \alpha\epsilon$ and
$\gamma^{2} - \delta\epsilon$ should all have the same sign. Since we will
normally assume that the stability parameters, $\delta$ and $\epsilon$, which
ensure that the potential is bounded below, are small, we will take 
$\beta\gamma > \alpha\epsilon$ and $\alpha\gamma > \beta\delta$, which 
implies that $\gamma^{2} > \delta\epsilon$.

The dynamics given by (\ref{Langevin}), (\ref{noise}) and (\ref{potential})
is equivalent to an overdamped particle moving on the potential surface 
shown in Fig \ref{Fig1}(a) and which is also acted upon by white noise. In
a typical realization, the particle will begin at the maximum and perform a 
random walk which is in the vicinity of the origin at early times, but which
eventually explores an ever larger region. Eventually, the particle gets far
enough from the origin that the deterministic dynamics, specified by $V$, 
begins to have a significant impact and the particle accelerates down towards
the saddles and the minima. In this paper our main concern is state selection:
we are not concerned with the approach the particle makes to a particular
minimum, since by this stage this minimum will almost certainly be the
selected state. For this not to be so, the particle would have to hop over
a barrier --- an extraordinarily rare event. In fact, as is clear from 
Fig \ref{Fig1}(a), as soon as the particle has passed the $x$ or $y$ 
coordinate of one of the saddle points, it has effectively chosen the final 
state and its subsequent motion is of little interest to us. We therefore
arrive at the conclusion that the saddle points are the major factor 
influencing state selection and by comparison the minima are of little 
consequence. This can be made more transparent if we imagine that $\delta$
and $\epsilon$ are very small, so that the minima (\ref{minima}) are now
at very large values of $x$ and $y$. By contrast, the positions of the
saddles have changed much less: in the limit where $\delta$ and $\epsilon$
go to zero their $x$ and $y$ coordinates tend to the finite values 
$\sqrt{\beta/\gamma}$ and $\sqrt{\alpha/\gamma}$, whereas the minima tend to
infinity. The topography now consists of long valleys, leading from the head
of the valley between two saddles, all the way down to a minima at the end 
of the valley. As soon as the particle enters the valley it is extremely 
unlikely to escape over the sides, and so extremely unlikely it will do 
anything else but move to the minima at the end of that particular valley. 
This makes it very clear that it is the saddles, and not the minima, that 
we should focus on if we wish to understand state selection.

These ideas are well illustrated by plotting out a few Monte Carlo 
trajectories. Such simulations are extremely simple to carry out, and later 
we will compare our analytic expression for the probability of a particular 
state being selected with the proportion of runs which ended up in that 
state. For the moment, however, we are interested in the nature of individual
trajectories. The five which are shown in Fig 2 are typical: the particle 
carries out Brownian motion about the origin, to a greater or lesser extent, 
and then selects a particular valley. Even after state selection, there may 
still be large deviations, but eventually the particle settles down near to 
the axis along which the valley runs.

\begin{figure}[tbh]
\begin{center}
\leavevmode
\hbox{\epsfxsize=0.8\columnwidth \epsfbox{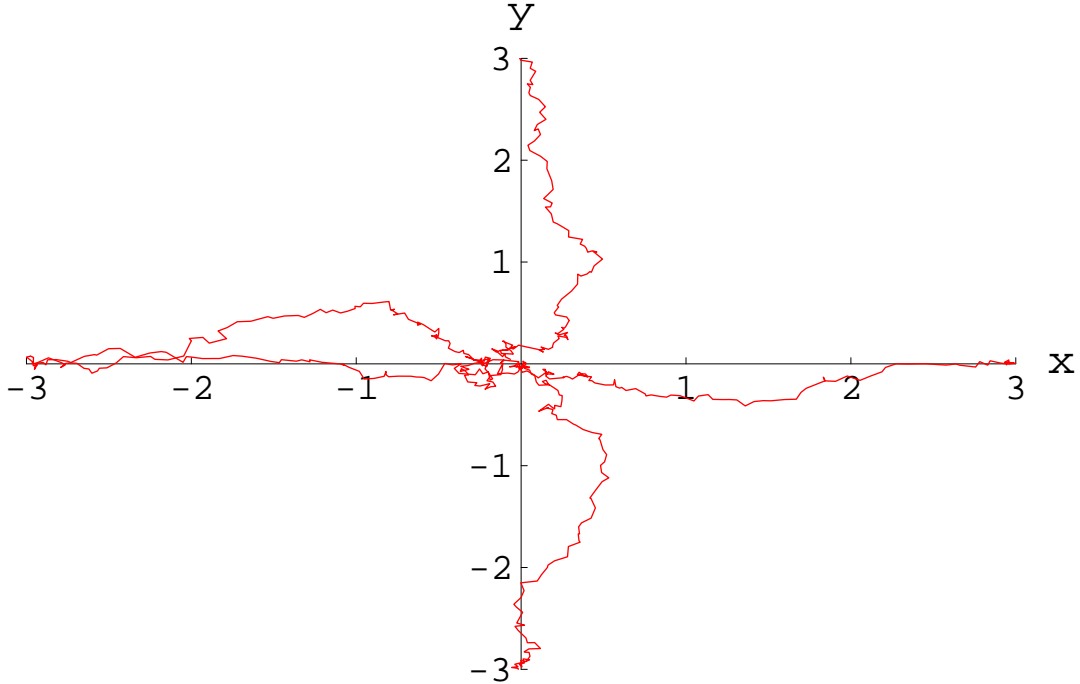}}
\end{center}
\caption{Typical trajectories for the full potential $V(x, y)$ with
$\alpha = \beta = \gamma = 1$, $\delta = \epsilon = 0.1$, and with noise 
strength $D=0.01$.}
\label{Fig2}
\end{figure}

\section{The Reduced Potential}

We have seen in the last section that the essential features of state 
selection become much clearer in the limit $\delta, \epsilon \rightarrow 0$,
when the saddle points are completely separated from the minima, which have
moved off to infinity. For the rest of the paper we work in this limit in
order to illustrate our method in the simplest possible way. We will call
the problem defined in Sec. II when the limit is taken, the reduced problem.
The reduced potential is then defined to be
\begin{equation}
V_{R}(x,y) = - \frac{\alpha}{2}x^{2} - \frac{\beta}{2}y^{2} + 
\frac{\gamma}{2}x^{2} y^{2}\,.
\label{red_potential}
\end{equation}
A plot of $V_{R}$ looks very similar to the plot of the full potential $V$
in Fig \ref{Fig1}(a), since this figure emphasizes the little-changed central
features of the potential. The main difference is that whereas in $V$ the 
valley floors start to ascend again at large $|x|$ and $|y|$, in $V_{R}$ they
keep descending. However, as was made clear in Sec II, this behavior is of no 
interest to us, and the fact that $V$ and $V_{R}$ differ in this way is
immaterial. 

From (\ref{red_potential}), we see that $V_{R}$ has only two types of 
extrema: (i) a maximum at the origin, and (ii) four saddle points at 
$(x,y) = (\pm X_{\rm min}, \pm Y_{\rm min})$ where
\begin{equation}
X_{\rm min} = \sqrt{\frac{\beta}{\gamma}}\ \ , \ \ 
Y_{\rm min} = \sqrt{\frac{\alpha}{\gamma}}\,.
\label{def_min}
\end{equation}
The reason for the subscript ``min'' is that the values given by 
(\ref{def_min}) are the smallest values of $|x|$ and $|y|$ at which we can 
assume that state selection has taken place. A simple stability analysis
near these saddles shows that the stable and unstable directions are at
an angle of $\pi/4$ to the axes. For example, in the positive quadrant, if
the particle approaches the saddle along the line of steepest descent 
$y=x$, it will have a tendency to move down the slopes which are to the
right and left, rather than carry on through the saddle and straight up
the incline ahead. 

Just as the plot of $V_{R}$ is similar to that of $V$ if we cut it off near
to the head of the valleys, so the Monte Carlo simulations are similar in
both cases if we do not follow the trajectories too deeply into the valleys. 
In Fig. 3, the range of $x$ and $y$ used to display the trajectories 
are smaller than those in Fig. 2, so that the central region is emphasized.
This makes the initial Brownian dynamics seem more obvious, but in reality the
nature of the trajectories before and during state selection are essentially
identical to those of the full problem shown in Fig. 2. 

\begin{figure}[tbh]
\begin{center}
\leavevmode
\hbox{\epsfxsize=0.8\columnwidth \epsfbox{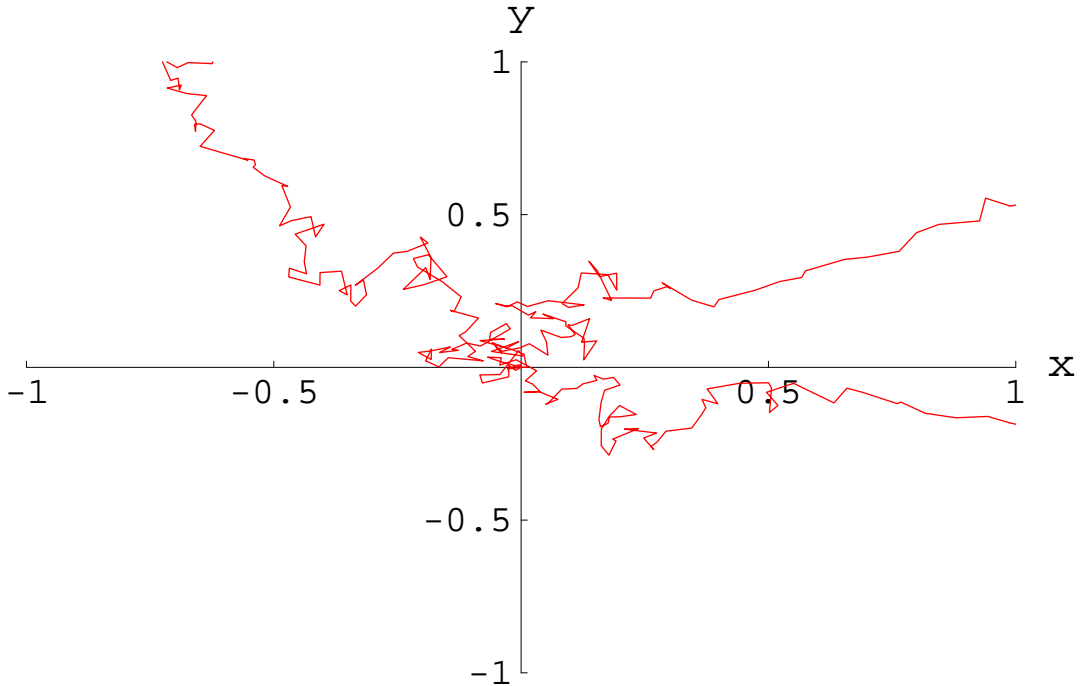}}
\end{center}
\caption{Typical trajectories for the reduced potential $V_{R}(x, y)$ with
$\alpha = \beta = \gamma = 1$, and with noise strength $D=0.01$.}
\label{Fig3}
\end{figure}

In the next section, we will develop a calculational scheme, based on the 
path-integral formulation of the stochastic process, to determine the 
probabilities of entering the $x$- or $y$-valleys as a function of the
model parameters $\alpha, \beta, \gamma$ and $D$. However, one might feel 
that such a sophisticated theory is not necessary: it should be possible to
obtain a satisfactory theory by constructing an approximation based on the
picture we have built up in this and the previous section. We will therefore
end this section by constructing an example of such a theory and show that 
it is unsatisfactory for a variety of reasons.

A simple theory of state selection might have the following ingredients:
\begin{itemize}
\item[1.]In the initial period the noise and linear growth of the modes 
dominate. Therefore, neglect the non-linear interaction between the modes 
(i.e. set $\gamma = 0$).
\item[2.]State selection will be specified in the following way. Define 
four sectors in the $xy$-plane by drawing lines through the origin and the
saddle points. The particle will select the state corresponding to a 
particular sector if it is in that sector for $x \approx X_{\rm min}$ and
$y \approx Y_{\rm min}$.
\end{itemize}

A calculation based on these two assumptions is given in Appendix A. Even 
within the framework that these provide, there turn out to be many
possible variants, each giving slightly different results. Leaving this 
aside for the moment, it is shown in Appendix A that in one of the simplest 
schemes along these lines, the probability of ending up in an $x$-valley is
(see Eqn. (\ref{naive_form3}))
\begin{equation}
N_{x} = \frac{1}{1 + \hat{D}^{\rho}}\,,
\label{naive1}
\end{equation}
where
\begin{equation}
\hat{D} \equiv \frac{\gamma D}{\alpha\beta} \ \ {\rm and} \ \
\rho = \frac{(\alpha - \beta)}{\alpha}\,.
\label{Dandrho}
\end{equation}
This has the correct qualitative behavior: when $\alpha = \beta$, that is,
$\rho = 0$, $N_{x} = 1/2$. As $\rho$ increases, $N_{x}$ increases towards
unity. A more careful calculation gives (Eqn. (\ref{naive_form4}))
\begin{equation}
N_{x} = \frac{2}{\pi} \tan^{-1} \hat{D}^{-\rho}\,.
\label{naive2}
\end{equation}
which again has the correct qualitative features. 

We now compare these two formulas with the results of Monte Carlo 
simulations. We took $\beta, \gamma$ and $D$ to be fixed in the simulations,
at values $1.0, 1.0$ and $0.01$ respectively, and varied $\alpha$ from $1.0$
to $3.0$ in steps of 0.2. For each value of $\alpha$ we performed a large
number of runs where the particle started at the origin and ended up 
in one or other of the valleys (for more details see Sec. VI). We then
simply counted the number of times that the particle ended up in a 
particular valley and expressed this as a fraction of the total number
of runs made. The results are shown in Fig. 4.

\begin{figure}[tbh]
\begin{center}
\leavevmode
\hbox{\epsfxsize=0.8\columnwidth \epsfbox{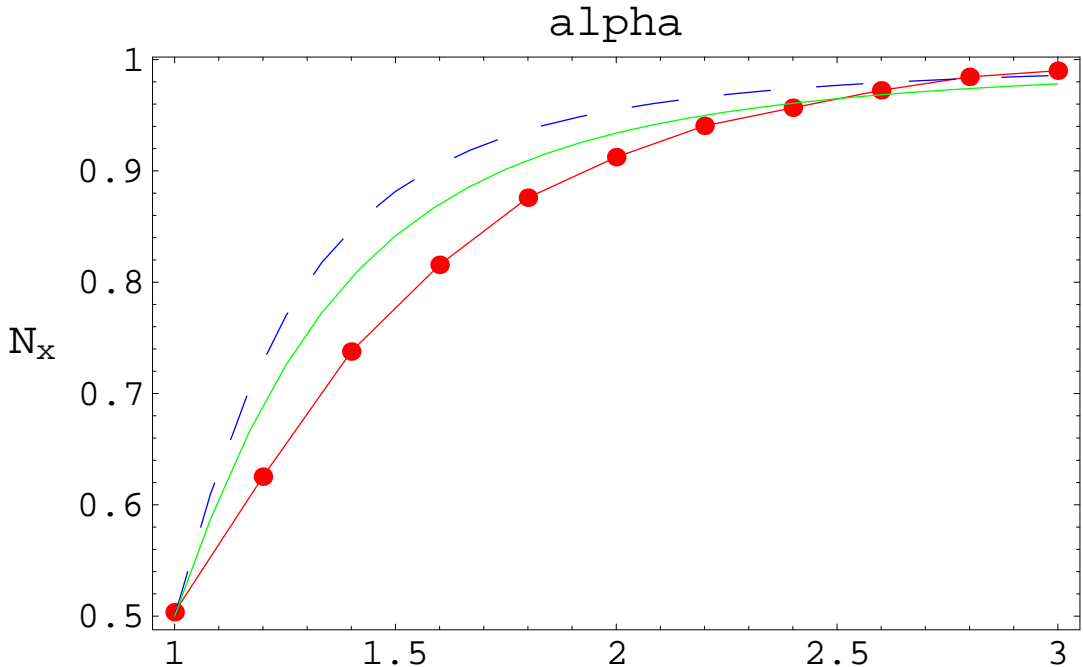}}
\end{center}
\caption{Probability of flowing into an $x$-valley as a function of $\alpha$,
with $\beta = 1$, $\gamma =1$ and $D = 0.01$. The dots show simulation 
results, the full curve is the result (\ref{naive1}) and the dashed curve
is result (\ref{naive2}).}
\label{Fig4}
\end{figure}

It is clear that the results of the naive approach based on assumptions 1 
and 2 above, worked out in Appendix A and given by (\ref{naive1}) and 
(\ref{naive2}) are not quantitatively correct. They mainly overestimate the 
probability of the particle ending up in the $x$-valley. In addition, Eqn.
(\ref{naive2}), which represents a more refined calculation, actually 
compares less favorably with the simulation results, than does (\ref{naive1}).
This clearly points to the unsatisfactory nature of this simple-minded 
scheme, as one would expect the more refined calculation to compare more 
favorably with the simulation results.

We do not wish to pursue techniques based on this approach further in this
paper. It was included specifically to show that there is a delicate 
interplay between the non-linearity and the noise in (\ref{x_eqn}) and 
(\ref{y_eqn}) and any attempt to incorporate one or the other of these 
using an unsystematic approximation scheme, as above, is likely to give 
disappointing results. We do not rule out being able to construct a 
particular scheme of this type which will give reasonable results: there 
are enough possible variants that it may be possible to interpolate 
between these by introducing free parameters which can then be fitted to 
the Monte Carlo data. However, such an {\it ad hoc} scheme is unnecessary, 
since we will show in the remainder of the paper that a systematic approach 
exists which gives simple formulas which are in good agreement with the 
Monte Carlo data. The method is based on the use of optimal paths, and 
so we begin with a brief review of the path-integral formulation of 
Langevin dynamics. 

\section{Optimal paths}

In this section we review the formulation of stochastic differential 
equations, such as (\ref{Langevin}), as functional integrals and obtain the 
dominant contribution to the conditional probability we wish to determine 
in the limit where the noise strength tends to zero. 

The conditional probability that the system is in the state $(x_{f},y_{f})$ 
at time $T$, given it was initially in the state $(0,0)$ at $t=0$ is
\begin{equation}
P(x_{f},y_{f},T|0,0,0) = \left\langle \delta \left( x_{f} - x(T) \right)
\delta \left( y_{f} - y(T) \right) \right\rangle_{\rm IC}\,,
\label{condprob_def}
\end{equation}
where IC denotes the initial condition $x(0) = 0$, $y(0) = 0$ on the 
stochastic process and $x(T)$ and $y(T)$ are solutions of (\ref{x_eqn}) and 
(\ref{y_eqn}). The average in (\ref{condprob_def}) is over Gaussian white 
noises $\eta_{x}(t)$ and $\eta_{y}(t)$ with zero mean and correlation 
function given by (\ref{noise}). In terms of functional integrals 
(\ref{condprob_def}) equals
\begin{equation}
{\cal C}\int_{\rm IC} D\eta_{x}D\eta_{y}\,
\delta\left( x_{f} - x_{\eta}(T) \right) \delta \left( y_{f} - 
y_{\eta}(T) \right)\,\exp\left\{ - \frac{1}{4D}\int^{T}_{0} dt
\left[ \eta^{2}_{x}(t) + \eta^{2}_{y}(t) \right] \right\}\,,
\label{functint_eta}
\end{equation}
where ${\cal C}$ is a normalization constant and the subscript $\eta$ on 
$x(T)$ and $y(T)$ is to emphasize that they depend on $\eta(T)$ through 
(\ref{x_eqn}) and (\ref{y_eqn}). Using these equations to perform a
functional change of variable from $(\eta_{x}, \eta_{y})$ to $(x,y)$ yields
\begin{equation}
{\cal C}\int_{\rm IC} Dx Dy\,J\,\delta\left( x_{f} - x(T) \right)
\delta\left( y_{f} - y(T) \right)\exp\left\{ - \frac{1}{4D}\int^{T}_{0} dt
\left[ \left( \dot{x} + \frac{\partial V}{\partial x} \right)^{2} +
\left( \dot{y} + \frac{\partial V}{\partial y} \right)^{2} \right] \right\}\,,
\label{functint_xy}
\end{equation}
where $J$ is the Jacobian of the transformation. Expressed as a path-integral
\begin{equation}
P(\vec{r}_{f},T|\vec{0},0) = {\cal C}
\int^{\vec{r}(T)=\vec{r}_{f}}_{\vec{r}(0)=\vec{0}} 
D\vec{r}\,J[\vec{r}]\,\exp\left\{ - S[\vec{r}]/D \right\}\,,
\label{pathint}
\end{equation}
where $\vec{r}=(x,y)$. The action $S[\vec{r}]$ and the Jacobian $J[\vec{r}]$ 
are functionals which are given by
\begin{equation}
S[x,y] = \frac{1}{4} \int^{T}_{0} dt \,\left[ \left( \dot{x} +
\frac{\partial V}{\partial x} \right)^{2} + \left( \dot{y} +
\frac{\partial V}{\partial y} \right)^{2} \right]\,,
\label{action}
\end{equation}
and\cite{Zinn}
\begin{equation}
J[x,y] = {\rm Det}\left[ \frac{\delta \vec{\eta}}{\delta \vec{r}} \right]
\propto \exp\left\{ \frac{1}{2} \int^{T}_{0} dt \left[ \frac{\partial^{2}V}
{\partial x^{2}} + \frac{\partial^{2}V}{\partial y^{2}} \right] \right\}\,.
\label{jacobian}
\end{equation}
For $D \rightarrow 0$, the path-integral (\ref{pathint}) is dominated by 
solutions of the Euler-Lagrange equations $\delta S/\delta \vec{r}(t) = 0$, 
which satisfy the boundary conditions $\vec{r}(0) = 0$ and 
$\vec{r}(T) = \vec{r}_{f}$. Let the solution of least action be denoted by 
$\vec{r}_{c}(t)$. Then writing
$\vec{r}(t) = \vec{r}_{c}(t) + \delta\vec{r}(t)$, we have
\begin{displaymath}
P(\vec{r}_{f},T|\vec{0},0) = 
{\cal C}\,\exp\left\{ - S(\vec{r}_{c})/D \right\}\,\int D\delta\vec{r}\,
J[\vec{r}_{c} + \delta\vec{r}] 
\end{displaymath}
\begin{equation}
\times \exp\left\{ - \frac{1}{D} \int^{T}_{0} dt' \int^{T}_{0}
dt'' \delta\vec{r}(t') \left[ \frac{1}{2} \left. \frac{\delta^{2} S}
{\delta\vec{r}(t')\delta\vec{r}(t'')} \right|_{\vec{r}=\vec{r}_{c}} \right]
\delta\vec{r}(t'') + O(\delta\vec{r})^{3} \right\}\,.
\label{Dexpansion}
\end{equation}
Scaling $\delta\vec{r}$ by $D^{1/2}$ and performing the Gaussian functional
integral yields
\begin{equation}
P(\vec{r}_{f},T|\vec{0},0) = {\cal C'}\,
\exp\left\{ - S(\vec{r}_{c})/D \right\}\,
J(\vec{r}_{c})\,{\rm Det}\left[ \left. \frac{\delta^{2}S}{\delta\vec{r}(t')
\delta\vec{r}(t'')} \right|_{\vec{r}=\vec{r}_{c}} \right]^{-1/2}
\left[ 1 + O(D) \right]\,.
\label{steepest_descent}
\end{equation}
The new overall constant, ${\cal C'}$, is immaterial since, as we have seen in
Sec. III, we are interested only in the probability of ending up in the 
$x$-valley or $y$-valley, and we normalize these probabilities according to
$N_{x} + N_{y} = 1$. We will therefore omit the overall constant from now on. 

While all the derivations in this section have been carried out for potential
systems with two degrees of freedom, it should be clear that they generalize 
in an obvious way to systems of more than two degrees of freedom and those 
where no potential exists\cite{Zinn}. We may summarize the result of 
performing the functional steepest descent on (\ref{pathint}) to 
next-to-leading order by
\begin{equation}
P(\vec{r}_{f},T|\vec{0},0) = P^{(1)}(\vec{r}_{f},T)
\exp\left\{ - P^{(0)}(\vec{r}_{f},T)/D \right\}\,\left[ 1 + O(D) \right] \,,
\label{starting_point}
\end{equation}
where the leading order contribution, $P^{(0)}$, is just the action of the 
optimal path $\vec{r}_{c}(t)$ and the next-to-leading order contribution is
\begin{equation}
P^{(1)}(\vec{r}_{f},T) = J(\vec{r}_{c})\,
\left( {\rm Det}L(\vec{r}_{c}) \right)^{-1/2}\,.
\label{P_1}
\end{equation}
Here $L$ is the matrix formed from the second order functional derivative of
the action functional evaluated at the optimal path:
\begin{equation}
L(\vec{r}_c) = \left. \frac{\delta^{2}S}{\delta\vec{r}(t')
\delta\vec{r}(t'')} \right|_{\vec{r}=\vec{r}_{c}}\,. 
\label{fluctmat_defn}
\end{equation}
The result (\ref{starting_point}) is the starting point for our method: if
we can determine the functions $P^{(0)}$ and $P^{(1)}$, then we will have
a form for the conditional probability valid when the noise is weak. This 
in turn will enable us to obtain a formula for the probability that either 
the $x$-valley or $y$-valley is selected, as a function of 
$\alpha, \beta, \gamma$ and $D$. We shall devote the rest of this section 
to the determination of $P^{(0)}$ and leave the calculation of $P^{(1)}$ 
to the next section.

The case of interest to us in this paper is when $V$ is the reduced potential
(\ref{red_potential}). The Euler-Lagrange equations for this problem are
\begin{eqnarray}
\ddot{x} & = & x \left( \alpha - \gamma y^{2} \right)^{2}
- 2\gamma x y^{2} \left( \beta - \gamma x^{2} \right)\ \ {\rm and}
\label{fullEL_x} \\
\ddot{y} & = & y \left( \beta - \gamma x^{2} \right)^{2}
- 2\gamma y x^{2} \left( \alpha - \gamma y^{2} \right)\,.
\label{fullEL_y}
\end{eqnarray}
It is important to realize that there are two distinct dynamics associated 
with the problem under consideration. The first is the stochastic dynamics
given by (\ref{Langevin}) with the potential (\ref{red_potential}). This
was our starting point, the basis of the intuitive discussion of the
dynamics given in Secs. II and III, and the dynamics of the Monte Carlo 
simulation. The second dynamics is the {\em deterministic} dynamics, given 
by (\ref{fullEL_x}) and (\ref{fullEL_y}), which describes the 
$D \rightarrow 0$ limit of the stochastic dynamics. They are quite different 
and it is important not to carry over intuition from one to the other without
careful consideration. From (\ref{action}),
\begin{equation}
S[\vec{r}] = \frac{1}{2} \int^{T}_{0} dt \left[ \frac{1}{2}\dot{\vec{r}}^{2}
- U(\vec{r}) \right] + \frac{1}{2}\int^{T}_{0} dt \frac{dV}{dt}\,,
\label{action_alt}
\end{equation}
where
\begin{equation}
U(\vec{r}) = - \frac{1}{2} \left( \frac{\partial V}{\partial x} \right)^{2}
-\frac{1}{2} \left( \frac{\partial V}{\partial y} \right)^{2}\,.
\label{potential_alt}
\end{equation}
Since the last term in (\ref{action_alt}) is a constant, and consequently 
gives zero variation, the Euler-Lagrange equations (\ref{fullEL_x}) and 
(\ref{fullEL_y}) correspond to classical mechanics in the potential 
\begin{equation}
U_{R}(x,y) = - \frac{1}{2} x^{2}(\alpha - \gamma y^{2})^{2}
- \frac{1}{2} y^{2}(\beta - \gamma x^{2})^{2}\,.
\label{red_altpot}
\end{equation}
When considering the optimal paths such as $\vec{r}_{c}(t)$, it is the
potential $U_{R}$, and not $V_{R}$, that is relevant. A plot of $U_{R}(x,y)$
is shown in Fig. 5.

\begin{figure}[tbh]
\begin{center}
\leavevmode
\hbox{\epsfxsize=0.8\columnwidth \epsfbox{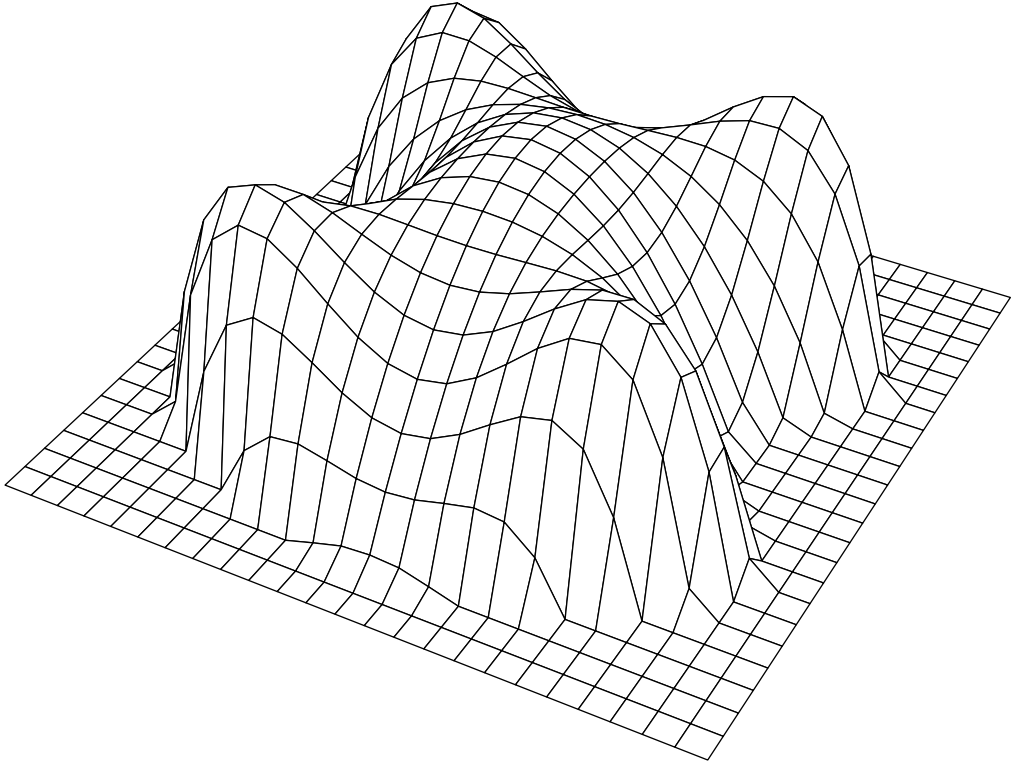}}
\end{center}
\caption{Illustrative plot of the reduced potential $U_{R}(x, y)$ 
with $\alpha = 2$, $\beta = \gamma = 1$.}
\label{Fig5}
\end{figure}

We will, for concreteness, focus on paths which end in the positive 
$x$-valley; paths which end in the other valleys are treated in exactly the 
same way. The boundary conditions on these paths are $x(0) = y(0) = 0$, 
$x(T) = x_{f}$ and $y(T) = y_{f}$, with $y_{f} \ll x_{f}$. We therefore 
simplify (\ref{fullEL_x}) and (\ref{fullEL_y}) by keeping only terms which 
are linear in $y$, to obtain to lowest order,
\begin{eqnarray}
\ddot{x} & = & \alpha^{2}x\ \ {\rm and}
\label{EL_x} \\
\ddot{y} & = & \left[ \beta^{2} -2(\alpha + \beta )\gamma x^{2}
+ \gamma^{2} x^{4} \right]\,y\,.
\label{EL_y}
\end{eqnarray}
In our earlier treatment of this problem\cite{I}, we restricted attention to 
end-points lying on the $x$-axis: $x_{f} = X$ and $y_{f} = 0$ (and 
correspondingly $y_{f} = Y$ and $x_{f} = 0$ for paths ending in the positive
$y$-valley) arguing that the relative contributions coming from paths 
ending on the axes should be approximately equal to the relative 
contributions coming from the paths ending anywhere in the valleys. In fact, 
as shown in Ref. \cite{I}, this turns out to be an excellent approximation, 
but it has the deficiency that $X$ is not determined. In this paper, we go 
beyond this approximation, which allows us to explore its validity, but 
also, as we will see, the precise value of $X$ within this improved 
treatment does not enter; the result is independent of $X$, provided that $X$
is large enough. The quantity $y_{f}$ has a different character to $X$: it 
will be integrated over at a later stage, when the probability flux through 
the valley is determined.

The solution of (\ref{EL_x}) which satisfies the boundary conditions 
$x(0) = 0$ and $x(T) = X$ is exactly the same as in Ref. \cite{I}, namely,
\begin{equation}
x_{c}(t) = X \frac{\sinh \alpha t}{\sinh \alpha T}\,.
\label{x_soln}
\end{equation}
The solution of (\ref{EL_y}) which satisfies the boundary conditions 
$y(0) = 0$ and $y(T) = y_{f}$ (we reserve the notation $Y$ for the final $y$
value for the end-point of paths going along the $y$-valleys) is discussed 
in Appendix B. There it is shown that an excellent approximation to the 
solution of this equation (indistinguishable from the numerical solution) 
can be found by first solving the equation for small $t$ and then for large 
$t$, and matching the two at some intermediate matching time $t_{m}$. 
Comparison with the numerical solution shows that the approximation remains 
good for a wide range of values of $t_{m}$ from about $0.1 T$ to $0.6 T$ or 
$0.7 T$. In order to be able to obtain a simple form for the solution, we 
have assumed that $T$ is large, in the sense that $e^{-\alpha T} \ll 1$ and 
$e^{-\beta T} \ll 1$. Using (\ref{smallt}) and (\ref{larget}), we can write 
the explicit analytic from as 
\begin{equation}
y_{c}(t) =
\left\{ \begin{array}{ll} 
\ \ \ \ \ \ \ 2y_{f}\,e^{-\beta T}\,e^{\gamma X^{2}/2\alpha}\,
\sinh\beta t, & \mbox{\ $0 \le t \le t_{m}$} \\ 
y_{f}\,e^{-\beta (T-t)}\exp\left[ \frac{\gamma X^{2}}{2\alpha}
\left\{ 1 - e^{-2\alpha(T - t)} \right\} \right], & \mbox{\  
$t_{m} \le t \le T$\, .}
\end{array} \right.
\label{y_soln}
\end{equation}
For self-consistency we need to check that $y_{c}(t)$ given by 
(\ref{y_soln}) is small, compared with $x_{c}(t)$, if we are to justify 
the linearization procedure which led to (\ref{EL_x}) and (\ref{EL_y}). 
From (\ref{y_soln}) it is straightforward to check that $y_{c}(t)$ has a 
maximum at $t=t'$, where $t'=T$, if $X < X_{\rm min}$ and 
$T - t' = \alpha^{-1}\,\ln (X/X_{\rm min})$, if $X > X_{\rm min}$. As 
discussed in Sec. III, $X_{\rm min}$ is the minimum value of $x$ at which 
state selection can be said to have been completed, and therefore we ask 
that $X \ge X_{\rm min}$. The maximum value of $y_{c}(t)$,
\begin{equation}
y_{c}(t') = y_{f}\,\left( \frac{X}{X_{\rm min}} \right)^{-\beta/\alpha}\,
\exp\left\{ \frac{\gamma}{2\alpha}\,\left( X^{2} - 
X^{2}_{\rm min} \right) \right\}\,,
\label{max_y}
\end{equation}
increases rapidly with $X$ and is already extremely large when
$X \sim 5X_{\rm min}$. It might therefore be tempting to argue that we need
to take $X$ to be much less than this value if the linearization procedure
is
to be valid. Since (\ref{max_y}) has its least value when $X=X_{\rm min}$,
this approximation would seem to be best for this choice of $X$, and in fact
this was the choice made in our earlier work\cite{I}.

In fact, there are a variety of reasons why choosing $X$ to be $X_{\rm min}$
is not suitable. First, the argument above, namely that if $X$ is too
large then the linearization procedure is invalid, is in fact incorrect. We
will find that the width of the distribution is so narrow in the $y$ 
direction that the range of integration for $y_{f}$ need only be
tiny --- in fact only going out to values of $y_{f}$ such that
$y_{f}^{2}\exp\left\{ \gamma X^{2}/\alpha \right\} \sim 1$. From (\ref{max_y})
we see that under these conditions $y_{c}(t)$ is indeed small. We will come
back to this point again in Secs. VI and VII.

The second reason is in fact more profound. Intuitively, we expect that
for large enough values of $X$ the probability flux through a given valley
should be independent of $X$. This is due to the fact that in the regime of
weak noise, once state selection has occurred there is no flux leakage out
of a valley. This implies that we should not have to make any choice for $X$
because our results should be independent of $X$ as long as $X$ is large
enough. In fact, this is precisely what we find, namely that the 
probability flux through the $x$-valley tends to an asymptotic value as $X$ 
tends to infinity. The remarkable cancellations that occur to produce an $X$ 
independent flux is a good indicator of the correctness of our calculational 
scheme. Again, we will come back to this point in Secs. VI and VII.

Finally, throughout our calculation we use the approximation that $T$ is 
large in the sense that $e^{-\alpha T} \ll 1$ and $e^{-\beta T} \ll 1$. This 
is apparently a problem, since we need the form of the distribution for 
all $T$ in order to perform the time integral of the probability flux in 
the calculation of the total flux through the valley. The way out of this 
impasse is to take $X$ sufficiently large that the probability current is 
essentially zero at small times, and only starts making appreciable 
contributions to the $T$ integral when $T$ is such that $e^{-\alpha T} \ll 1$
and $e^{-\beta T} \ll 1$. It is not clear how large $X$ will have to be, 
but as discussed above, the probability flux through the $x$-valley tends 
to an asymptotic value as $X$ grows, so we are assured of always finding a 
value of $X$ for which the approximation that $T$ is large is valid.

At first sight, the large $T$ approximation might appear to be problematic
because it ignores the shorter state selection time scales $\alpha^{-1}$
and $\beta^{-1}$. However, this is not the case. The reason is that the
integrals over time that are performed in the calculation of the action 
and Jacobian prefactor include these earlier times. In other words, the 
entire classical path, in particular the spatial region near the origin, 
is included in the calculation.

After these rather technical asides, it is worthwhile summarizing what we 
have deduced about the equation for the optimal path which leaves $(0,0)$ at 
$t=0$ and arrives at $(X, y_{f})$ at time $t=T$. It is given by (\ref{x_soln})
and (\ref{y_soln}) under the assumptions that (i) $T$ is such that
$e^{-\alpha T} \ll 1$ and $e^{-\beta T} \ll 1$, (ii) $X$ is larger than a 
few $X_{\rm min}$, and (iii) $t_{m}$ and $(T-t_{m})$ are large (we will
later find that we require that 
$e^{-2\alpha t_{m}} \ll 1, e^{-2\beta t_{m}} \ll 1, 
e^{-2\alpha (T-t_{m})} \ll 1$ and $e^{-2\beta (T-t_{m})} \ll 1$).

We are now in a position to calculate the leading order contribution,
$P_{0}(X, y_{f}, T)$, to the conditional probability distribution 
(\ref{starting_point}), by finding the action of the optimal path given by 
(\ref{x_soln}) and (\ref{y_soln}). We begin by noting that, from
(\ref{action_alt}), the classical paths are solutions of 
$\ddot{x} = - \partial U/\partial x$ and $\ddot{y} = - \partial U/\partial y$.
Multiplying the first equation by $\dot{x}$, the second one by $\dot{y}$,
adding them and integrating the result, gives 
$\frac{1}{2}(\dot{x}^{2}_{c} + \dot{y}^{2}_{c} ) + U = E$, a constant. Thus
another form for the action of classical paths is
\begin{equation}
S_{c} = \frac{1}{2} \int^{T}_{0} dt\,\left[ \dot{x}^{2}_{c} + 
\dot{y}^{2}_{c} \right]
- \frac{1}{2} E T + \frac{1}{2} \left[ V \right]^{T}_{0}\,.
\label{class_action}
\end{equation}
Details of the calculation are given in Appendix B. From (\ref{cla_action})
we find
\begin{equation}
S_{c} = \frac{\alpha X^{2}}{4}\,\left( {\rm coth}\alpha T - 1 \right)
+ \frac{1}{2}\,y^{2}_{f}\,S_{y}\,,
\label{cl_action}
\end{equation}
where
\begin{equation}
S_{y} = \frac{1}{2 \alpha}\,\left( \frac{\alpha}
{\gamma X^{2}} \right)^{\beta/\alpha}\,\exp\left\{ \frac{\gamma X^{2}}
{\alpha} \right\}\,
\int^{\gamma X^{2}/\alpha}_{0} dz\,z^{(\beta/\alpha) - 1}\, e^{-z}\,
(\beta - \alpha z)^{2} + \frac{1}{2} [\gamma X^{2} - \beta]\,.
\label{Sy}
\end{equation}
The two terms in (\ref{cl_action}) have different characters: as remarked 
earlier, we will eventually integrate over $y_{f}$ to obtain the probability 
flux through the valley, but the first term will remain, giving a leading 
order contribution of $\exp{-(\alpha X^{2}/4 D)({\rm coth}\alpha T - 1)}$
to (\ref{starting_point}). 

When performing similar calculations to find the escape rate from one
metastable state to another, the leading order result $e^{-\Delta V/D}$
gives a reasonable estimate, and the need to go to next order and to 
calculate the prefactor only arises if increased accuracy is required.
The situation is very different in the present case of the decay from 
an unstable state. As mentioned already, we expect that the main contribution
to the probability flux through the valley will occur when 
$T \sim (2\alpha)^{-1}\,\ln (\alpha X^{2}/D)$. The conditional probability  
distribution (\ref{starting_point}) is peaked around a value of $T$ of this
order because of a balance between the leading (classical) term and the
prefactor (a fluctuational) term. For this reason, the calculation of the 
prefactor in this problem, unlike in escape problems, is vital if the 
essential structure of the state selection probabilities is to be captured. 
We therefore turn to the calculation of this quantity.

\section{Calculation of the prefactor}

In this section we will calculate the next to leading order contribution in
(\ref{starting_point}). From (\ref{P_1}) we see that it consists of two 
distinct contributions: from the Jacobian $J[\vec{r}]$ evaluated at the 
optimal path, and from the fluctuations around the optimal path which give
rise to the determinant of the second functional derivative of the action
(\ref{fluctmat_defn}). We begin by evaluation of this determinant.
 
Starting from (\ref{action_alt}), and using the notation 
(\ref{fluctmat_defn}), we obtain
\begin{eqnarray}
\def\linie{\vrule height 17pt depth 5pt}
\def\back{\noalign{\vskip-3pt}}
\lineskip=0pt
L(\vec{r}_{c}) = \left[
\matrix{ &   & \linie & &   & & \cr
         & -\frac{d^{2}}{dt^{2}} + \alpha^{2} & \linie & & -4\gamma x_{c}
y_{c} \left[ \alpha + \beta -\gamma x_{c}^{2} \right] & &  \cr
       \back
         &   & \linie & &   & & \cr
           \noalign{\hrule}
         &   & \linie & &   & &  \cr
       \back
         & -4\gamma x_{c} y_{c} \left[ \alpha + \beta - 
\gamma x_{c}^{2} \right] & \linie & & -\frac{d^{2}}{dt^{2}} 
+ \left( \beta - \gamma x_{c}^{2} \right)^{2} - 
2\alpha \gamma x_{c}^{2} & &  \cr
       \back
         &   & \linie & &   & &  \cr
                } \right]\,,
\label{matrix}
\end{eqnarray}
where, in line with our previous assumption, we have neglected terms which 
are $O(y_{c}^{2})$ down on the terms shown in (\ref{matrix}). We have
omitted any time dependence in (\ref{matrix}) for the sake of clarity, but
it should be remembered that the classical solutions $x_{c}$ and $y_{c}$ are 
functions of $t$ and the matrix is multiplied by an overall factor of
$\delta ( t - t' )$. The diagonal entries are as in the $y_{c} = 0$ 
case\cite{I} --- only the off-diagonal entries are different. However, it
is easy enough to see that the eigenvalues of (\ref{matrix}) are even in
$y_{c}$, and since the matrix with $y_{c}=0$ has no zero eigenvalues, the
off-diagonal entries only provide $O(y^{2}_{c})$ corrections to the 
eigenvalues of the matrix with $y_{c}=0$. Therefore neglecting these terms 
as before, we conclude that within the approximation we have adopted
in this paper, we may set the off-diagonal entries in (\ref{matrix}) to
zero. From (\ref{fluctmat_defn}) and (\ref{matrix}) we now find
\begin{equation}
{\rm Det}L(\vec{r}_{c}) = {\rm Det}L_{x}\,{\rm Det}L_{y}\,,
\label{det_prod}
\end{equation}
where $L_{x}$ and $L_{y}$ represent fluctuations in $x$ and $y$ 
respectively:
\begin{eqnarray}
L_{x} & = & - \frac{d^{2}}{dt^{2}} + \alpha^{2}\,,
\label{Lx_defn} \\
L_{y} & = & - \frac{d^{2}}{dt^{2}} + \left( \beta - 
\gamma x^{2}_{c} \right)^{2} - 2\alpha\gamma x^{2}_{c}\,.
\label{Ly_defn}
\end{eqnarray}
To evaluate the determinants of the operators (\ref{Lx_defn}) and 
(\ref{Ly_defn}) we make use of a well-known formula for such 
determinants\cite{dets}. Let $L$ be an operator of the type 
$-d^{2}/dt^{2} + \phi(t)$ and suppose that the eigenfunctions of $L$ are 
required to vanish at the boundaries at $t=0$ and $t=T$, as in the case of 
interest to us here. Now let $h_{1}(t)$ and $h_{2}(t)$ be two independent 
solutions of the homogeneous equation $Lh=0$. Then
\begin{equation}
{\rm Det}L \propto \frac{h_{1}(0)h_{2}(T) - h_{2}(0)h_{1}(T)} 
{h_{1}(0)\dot{h}_{2}(0) - h_{2}(0)\dot{h}_{1}(0)}\,.
\label{det_formula_1} 
\end{equation}
The constant of proportionality in (\ref{det_formula_1}) may be omitted for 
the same reason the overall constant was omitted in (\ref{steepest_descent}):
the probability of ending up in the positive $x$-valley will be normalized
by the sum of the probabilities of ending up in the $x$- and $y$-valleys. 
Furthermore, if $h_{2}$ is the solution which vanishes at $t=0$, then the
formula for ${\rm Det}L$ involves only this solution:
\begin{equation}
{\rm Det}L = \frac{h_{2}(T)}{\dot{h}_{2}(0)}\,.
\label{det_formula_2} 
\end{equation}
The solution of $L_{x}h=0$ which satisfies $h(0)=0$ is 
$h_{2}(t)= F\sinh\alpha t$, where $F$ is a constant. Therefore
\begin{equation}
{\rm Det}L_{x} = \frac{\sinh\alpha T}{\alpha}\,.
\label{Lx}
\end{equation}
To find ${\rm Det}L_{y}$ we need only note that the solutions of the 
homogeneous equation $L_{y}h=0$ are also the solutions of the classical
equation (\ref{EL_y}). But the solution which vanishes at $t=0$ has 
already been found in Appendix B: it is given by (\ref{smallt}) and
(\ref{larget}),
\begin{equation}
h_{2}(t) =
\left\{ \begin{array}{ll} 
\ \ \ \ \ \ \ G \sinh\beta t, & \mbox{\ $0 \le t \le t_{m}$} \\ 
\frac{G}{2}\,e^{\beta t}\exp\left[ - \frac{\gamma X^{2}}{2\alpha}\,
e^{-2\alpha(T - t)} \right], & \mbox{\ $t_{m} \le t \le T$\, ,}
\end{array} \right.
\label{h2}
\end{equation}
where $G$ is a constant. Since $\dot{h}_{2}(0)=\beta G$, use of  
(\ref{det_formula_2}) yields,
\begin{equation}
{\rm Det}L_{y} = \frac{1}{2\beta}\,e^{\beta T}\,
\exp\left\{ - \frac{\gamma X^{2}}{2\alpha} \right\} \,.
\label{Ly}
\end{equation}

It only remains to calculate the Jacobian at the optimal path. From 
(\ref{jacobian}) this is given by
\begin{equation}
J(\vec{r}_{c}) = \exp\left\{ \frac{1}{2} \int^{T}_{0} dt 
\left[ \left( -\alpha + \gamma y^{2}_{c}(t) \right) + 
\left( - \beta + \gamma x^{2}_{c}(t) \right) \right] \right\}\,.
\label{class_jacobian}
\end{equation}
Neglecting the $O(y_{c}^{2})$ term as before, one finds that 
$J(\vec{r}_{c}) = J_{x} J_{y}$ where
\begin{eqnarray}
J_{x} & = & \exp\left\{ - \frac{1}{2} \alpha T \right\}\,,
\label{Jx} \\
J_{y} & = & \exp\left\{ - \frac{1}{2} \beta T + \frac{\gamma X^{2}}{4\alpha}\,
\left[ \frac{\sinh 2\alpha T - 2\alpha T}
{2\sinh^{2}\alpha T} \right] \right\} \nonumber \\
& \approx & \exp\left\{ - \frac{1}{2} \beta T + \frac{\gamma X^{2}}
{4\alpha} \right\}\,,
\label{Jy}
\end{eqnarray}
since we are assuming that $e^{-\alpha T} \ll 1$. Since both the determinants
and the Jacobians factorize into contributions associated with the $x$ 
coordinate and contributions associated with the $y$ coordinate, we may
summarize the results of this section so far as
\begin{eqnarray}
\frac{J_{x}}{\sqrt{{\rm Det}L_{x}}} & = & (2\alpha)^{1/2}\,e^{-\alpha T}\,,
\label{final_sec_x} \\
\frac{J_{y}}{\sqrt{{\rm Det}L_{y}}} & = & (2\beta)^{1/2}\,e^{-\beta T}\,
\exp \left\{ \frac{\gamma X^{2}}{2\alpha} \right\}\,.
\label{final_sec_y}
\end{eqnarray}
It is straightforward to check that these results agree with those of 
Ref. \cite{I} under the assumptions that $e^{-\alpha T} \ll 1$ and
$e^{-\beta T} \ll 1$. While neglecting corrections to the leading behavior 
of the prefactor which are of the type $e^{-2\alpha T}$ or $e^{-2\beta T}$ 
is in line with this large $T$ approximation, we will see later that the 
final integration over $T$ will also provide {\it a posteriori} 
justification for the neglect of these terms.  

From (\ref{starting_point}), (\ref{cl_action}), (\ref{final_sec_x}) and 
(\ref{final_sec_y}), we can now write down the conditional probability for 
the system to be at $\vec{r}_{f} = (X, y_{f})$ at time $T$, given that it 
was at the origin at time $t=0$, as
\begin{eqnarray}
P(\vec{r}_{f}, T|\vec{0}, 0) & = & \left( 4\alpha \beta \right)^{1/2}\,
\exp\left\{ - (\alpha + \beta )T + \frac{\gamma X^{2}}{2\alpha} \right\} 
\nonumber \\
& \times & \exp\left\{- \frac{\alpha X^{2}}{4 D}\,
\left( {\rm coth}\alpha T - 1 \right) \right\} 
\exp\left\{ - \frac{y^{2}_{f}}{2D}\,S_{y} \right\}\,\left[ 1 + O(D) \right]\,.
\label{final_P}
\end{eqnarray}
 
As we discussed in detail in Secs. II and III, when using the modified 
potential (\ref{red_potential}) the question is not what are the paths of 
least action from the origin to the minima of the potential, since those
minima cease to exist when the limit $\delta, \epsilon \rightarrow 0$ is
taken. Instead, the question is what is the relative flux through one 
valley compared to the other. To calculate the flux, it is not the 
conditional probability distribution (\ref{final_P}) that we need to know,
but the probability current $\vec{\cal J} = ({\cal J}_{x}, {\cal J}_{y})$.
This current is related to the conditional probability distribution 
through the Fokker-Planck equation\cite{Risken}
\begin{equation}
\frac{\partial P}{\partial t} + {\rm div} \vec{\cal J} = 0\,,
\label{FP}
\end{equation}
where
\begin{equation}
\vec{\cal J} = - \left( \nabla V \right) P - D \nabla P\,.
\label{current_def}
\end{equation}

In order to find $\vec{\cal J}$ from $P$ given by (\ref{final_P}), we first
note from (\ref{starting_point}) that
\begin{equation}
D \nabla P = - \left( \nabla P^{(0)} \right) P \left[ 1 + O(D) \right]\,.
\label{part_current}
\end{equation}   
Therefore we only need to differentiate $V$ and $P^{(0)}$ with respect to
$X$ and $y_{f}$ in order to find $\vec{\cal J}$. Carrying this out we 
obtain
\begin{eqnarray}
{\cal J}_{x} & = & \left( \frac{\alpha}{2} X \left[ {\rm coth} (\alpha T)
-1 \right] + \alpha X - \gamma X y_{f}^{2} \right) P 
\label{full_Jx} \\
{\cal J}_{y} & = & \left( S_{y} y_{f} + \beta y_{f} - \gamma y_{f} 
X^{2} \right) P \,.
\label{full_Jy}
\end{eqnarray}
In the next section we will calculate the flux through the positive 
$x$-valley. This will involve only the component of $\vec{\cal J}$ normal
to the line $x=X$. Therefore, the $y$-component of the current, 
${\cal J}_{y}$, will not contribute and need not be considered any further;
the entire contribution to the flux will come from integrating 
${\cal J}_{x}$, given by (\ref{full_Jx}), over $y_{f}$. By virtue of the 
$y_{f}^{2}$ term in the exponential in (\ref{final_P}) this will be a 
Gaussian integral. While we would naturally neglect the $O(y_{f}^{2})$ term 
in (\ref{full_Jx}) in line with previous approximations in this section, we 
now see that it would in any case give a contribution of order $D$ once the 
$y_{f}$ integral is performed. This is a further justification for ignoring 
such terms. Finally, we have been neglecting $e^{-2\alpha T}$ type 
corrections in the prefactor, which means that we should replace 
${\rm coth}\alpha T$ by $1$. These approximations lead to the result 
${\cal J}_{x} = \alpha X P$ and so we find from (\ref{final_P})
\begin{eqnarray}
{\cal J}_{x}(X, y_{f}, T) & = & 2 \alpha (\alpha \beta )^{1/2}\,X\,
\exp\left\{ - (\alpha + \beta )T + \frac{\gamma X^{2}}{2\alpha} \right\} 
\nonumber \\
& \times & \exp\left\{- \frac{\alpha X^{2}}{4 D}\,
\left( {\rm coth}\alpha T - 1 \right) \right\} 
\exp\left\{ - \frac{y^{2}_{f}}{2D}\,S_{y} \right\}\,\left[ 1 + O(D) \right]\,.
\label{final_Jx}
\end{eqnarray}
This is the key result of this section. We need now only use it to calculate
the total flux through the positive $x$-valley at $x=X$ and compare it with 
the analogous quantity through the positive $y$-valley. This will give us 
the relative probability of the $x$-mode being selected.

\section{Results}

The idea underlying the method we use to calculate the relative probability 
of one of the states being selected is most easily understood by first 
describing the analogous procedure used when carrying out Monte Carlo 
simulations. As was briefly alluded to in Sec. III, for each run the 
particle starts at the origin and subsequently follows the Langevin dynamics 
of the reduced problem until it reaches $x=\pm X$ in one of the
$x$-valleys or $y=\pm Y$ in one of the $y$-valleys. We used $X=5X_{\rm min}$ 
and $Y=5Y_{\rm min}$ to be certain that state selection had occurred. At 
this point that particular run ceases. If the particle has ended up in a 
$y$-valley, for instance, 1 is added to the total number of runs which have 
selected the $y$-mode. Another run is initiated and the result of that is 
added to the totals. After a large number of runs the proportions selecting 
the $x$-mode and $y$-mode are used to calculate $N_{x}$ and $N_{y}$, the 
relative probabilities of these states being selected. 

Let us focus on the runs ending up in the positive $x$-valley, exactly as 
we have been doing in the analytic treatment in Secs. IV and V. For each of
these runs the final value of $y$ --- called $y_{f}$ in the analytic
treatment above --- will vary. We would expect most of the runs to end near 
to the $x$-axis, with off-axis end-points becoming less and less common as 
we move away from the axis. This is reflected in the $y_{f}$ dependence of
${\cal J}_{x}$ given by (\ref{final_Jx}). Just as in the Monte Carlo 
simulation, where we add up all contributions with differing final 
$y$-coordinate at $x=X$, so we need to integrate (\ref{final_Jx}) over all 
$y_{f}$ at $x=X$. Similarly, just as in the Monte Carlo simulation, where we 
add up all of the contributions, no matter how long they took to get to the 
end-point, so in the analytic treatment we have to integrate over all $T$ to 
obtain the total flux. Therefore we need to calculate
\begin{equation}
{\cal F}_{x} (X)= \int^{\infty}_{0} dT\, \int^{\infty}_{-\infty} dy_{f}\,
{\cal J}_{x}(X, y_{f}, T)\,,
\label{Fx_defn}
\end{equation}
in order to calculate the state selection probabilities. 

While the above justification for (\ref{Fx_defn}) as the quantity which we
need to calculate seems intuitively plausible, it is worthwhile formally
proving this. We first need to specify what ``ending up in the positive 
$x$-valley'' means in terms of a mathematical expression. Since, in the 
limit $T \rightarrow \infty$, trajectories will have entered one of the 
four valleys, we define the probability that it has entered the positive 
$x$-valley as
\begin{equation}
{\rm Prob}(+ x{\rm -valley}) = \lim_{T \rightarrow \infty}\,
\int^{\infty}_{-\infty} dy_{f}\,\int^{\infty}_{X} 
dX\,P(X, y_{f}, T| \vec{0}, 0)\,.
\label{fund_defn}
\end{equation} 
Here the positive $x$-valley is defined as the entire potential surface 
to the right of the line $x=X$. On the other hand from the continuity 
equation (\ref{FP})
\begin{eqnarray}
\lim_{T \rightarrow \infty}\,P(X, y_{f}, T| \vec{0}, 0) - 
P(X, y_{f}, 0| \vec{0}, 0) & = &
\int^{\infty}_{0} \frac{\partial }{\partial T}\,P(X, y_{f}, T| \vec{0}, 0)\,
dT \nonumber \\ \nonumber \\
& = & - \int^{\infty}_{0} {\rm div} \vec{\cal J}\,dT\,.
\label{intermed_step1}
\end{eqnarray}
Integrating (\ref{intermed_step1}) over the ``volume'' 
$\{ x \ge X, -\infty < y_{f} < \infty \}$, using that fact that 
at $T=0$ the end-point is at the origin and making use of the divergence 
theorem, gives
\begin{equation}
\lim_{T \rightarrow \infty}\,\int^{\infty}_{-\infty} dy_{f}\,
\int^{\infty}_{X} dX\,P(X, y_{f}, T| \vec{0}, 0) =
- \int^{\infty}_{0} dT\,\int_{S} \vec{\cal J}.d\vec{S}\,.
\label{intermed_step2}
\end{equation}
The ``surface'' integral on the right-hand side of (\ref{intermed_step2})
only gives a contribution at $x=X$, where $d\vec{S}$ is in the direction of 
the outward normal, i.e. in the negative $x$ direction. Thus
$\vec{\cal J}.d\vec{S} = - {\cal J}_{x} dy_{f}$ and so the required 
probability (\ref{fund_defn}) is equal to (\ref{Fx_defn}). 
 
Substituting (\ref{final_Jx}) into (\ref{Fx_defn}) and performing the 
$y_{f}$ integration yields 
\begin{eqnarray}
{\cal F}_{x}(X) & = & 2 \alpha (\alpha \beta )^{1/2}\,X\,\left( \frac{2\pi D}
{S_{y}} \right)^{1/2}\,\exp \left\{ \frac{\gamma X^{2}}{2\alpha} \right\}\,
\int^{\infty}_{0} dT\,e^{-(\alpha + \beta )T} \nonumber \\
& \times & \exp\left\{- \frac{\alpha X^{2}}{4 D}\,
\left( {\rm coth}\alpha T - 1 \right) \right\} \left[ 1 + O(D) \right]\,. 
\label{Fx}
\end{eqnarray}
As remarked in Sec. V, any $O(y_{f}^{2})$ corrections in the prefactor 
give $O(D)$ corrections to (\ref{Fx}), as do $O(y_{f}^{4})$ corrections to
the action, which justifies their omission. The final integral over $y_{f}$ 
also provides justification for the linear approximation. An examination 
of ${\cal J}_{x}(X, y_{f}, T)$ shows that this function is effectively 
non-zero only for small values of $|y_{f}|$, which get still smaller 
as $X$ increases, so that unless 
$|y_{f}| \alt \exp\left\{ - \gamma X^{2}/2\alpha \right\}$, the flux is 
essentially zero. Thus either the conditional probability distribution or 
the conditional probability current resemble a thin wafer centered on the 
$x$-axis when plotted at fixed $T$. Moreover, this wafer gets very much 
thinner with increasing $X$. This means that the limits on the $y_{f}$ 
integral in (\ref{Fx_defn}) are effectively  
$ \pm \exp\left\{ - \gamma X^{2}/2\alpha \right\}$, which is tiny for large 
$X$. It is more instructive to change variables from $y_{f}$ to
$y^{*}_{f} = y_{f}\,\exp\left\{ \gamma X^{2}/2\alpha \right\}$, so that 
the integral now has the range $(-1, 1)$. But, as discussed in Sec. IV, this
means that (\ref{max_y}) now reads
\begin{equation}
y_{c}(t') = y^{*}_{f}\,\left( \frac{X}{X_{\rm min}} \right)^{-\beta/\alpha}\,
\exp\left\{ - \frac{\gamma X^{2}_{\rm min}}{2\alpha}\, \right\}\,,
\label{max_y_star}
\end{equation}
which has a magnitude less than one for $-1 < y^{*}_{f} < 1$. Thus the 
linearization procedure is justified. In some sense, the variable $y^{*}_{f}$
is more appropriate in this problem than $y_{f}$ itself.

An examination of the integrand in (\ref{Fx}) shows that is has a maximum
when $T=T^{*} \equiv (2\alpha)^{-1}\ln(\alpha^{2}X^{2}/[(\alpha + \beta)D])$,
which is slightly different to the value mentioned in Sec. IV. This is 
because that value came from the estimates of Sec. III and Appendix A, 
which used the value for the maximum which is found in the analogous 
one-dimensional stochastic process in $x(t)$. The more refined calculation
we have been discussing here has an additional $e^{-\beta T}$ contribution 
coming from fluctuations in $y(t)$ and from $J_{y}$. It is the occurrence 
of this term in the integrand of (\ref{Fx}) which changes the value at 
which the maximum occurs by this very slight amount. From (\ref{Fx}) we can 
also check the conjecture made in Sec. IV, that if $X$ is greater than a 
few $X_{\rm min}$, then even the tails of the integrand will be at values 
of $T$ which are large enough for the approximations made in this paper to 
be valid. A numerical investigation of the integrand shows that it has 
typically already fallen by two orders of magnitude from its maximum value 
for $T$ satisfying $e^{-\alpha T}, e^{-\beta T} \ll 1$, when $X$ is larger 
than a few $X_{\rm min}$. 

To perform the integration in (\ref{Fx}) we (i) replace 
$({\rm coth}\alpha T - 1)$ by its large $T$ form, $2e^{-2\alpha T}$ 
(justified below), and (ii) change variables to 
$\xi = (\alpha X^{2}/2 D) e^{-2\alpha T}$. Then the integral becomes
\begin{equation}
\frac{1}{2\alpha}\,\left( \frac{2 D}
{\alpha X^{2}} \right)^{(\alpha + \beta)/2\alpha}\,\int^{\Xi}_{0} d\xi\,
\xi^{[(\alpha + \beta)/2\alpha] - 1}\,e^{-\xi}\,,
\label{time_int}
\end{equation}
where $\Xi = (\alpha X^{2})/2 D$. The integral is an incomplete 
gamma function: it equals $\Gamma ( [\alpha + \beta]/2\alpha )$ plus 
exponentially small corrections in $D$ coming from the large, but finite,
upper limit $\Xi$\cite{AS}. Neglecting these exponentially small terms 
we therefore obtain for the total flux through the positive $x$-valley at 
$X$
\begin{equation}
{\cal F}_{x}(X) = (\alpha \beta )^{1/2}\,X\,\left( \frac{2\pi D}
{S_{y}} \right)^{1/2}\,\exp\left\{ \frac{\gamma X^{2}}{2\alpha} \right\}\,
\left( \frac{2 D}{\alpha X^{2}} \right)^{(\alpha + \beta)/2\alpha}\,
\Gamma \left( \frac{\alpha + \beta}{2\alpha} \right) \left[ 1 + 
O(D) \right]\,.
\label{final_FxX}
\end{equation}
By using the approximation $({\rm coth}\alpha T - 1) \approx 2e^{-2\alpha T}$,
terms of order $D^{-1} e^{-2(n+1)\alpha T}$, $(n=1,\ldots)$ were neglected,
but these are of the order of $D^{n} \xi^{n+1}$ and give contributions which 
are $O(D)$ down on the leading term (\ref{final_FxX}). Similarly, any 
corrections to the prefactor of order $e^{-\alpha T}$ or $e^{-\beta T}$ 
would also give contributions which are down compared to (\ref{final_FxX}).
Once again, we see that these final integrals justify approximations which
we made earlier. 

The total flux given by (\ref{final_FxX}) depends on $X$, which is still 
undetermined, apart from the requirement that it should be greater than a 
few $X_{\rm min}$. However, if we investigate the $X$ dependence of the
expression in (\ref{final_FxX}), we find that as $X$ increases from 
$X_{\rm min}$, it first decreases until $X$ equals two or three times 
$X_{\rm min}$, where it reaches a constant value and subsequently remains 
at this value as $X \rightarrow \infty$. Of course, this constancy of the
flux is exactly what we would expect: once the system has selected the 
positive $x$-valley, it remains in that state and so the flux should be 
conserved within the valley, i.e. be the same for all $X$. We will denote 
this constant value simply as ${\cal F}_{x}$; it is given by the 
$X \rightarrow \infty$ limit of (\ref{final_FxX}):
\begin{eqnarray}
{\cal F}_{x} & = & 2D\,(2\pi)^{1/2}\,(\alpha \beta)^{1/2}\,
\left( \frac{2\gamma D}{\alpha^{2}} \right)^{\beta/2\alpha}\,
\Gamma \left( \frac{\alpha + \beta}{2\alpha} \right) \nonumber \\
& \times &  \left\{ \int^{\infty}_{0} dz\,z^{(\beta/\alpha) - 1}\, e^{-z}\,
(\beta - \alpha z)^{2} \right\}^{-1/2}\,\left[ 1 + O(D) \right]\,.
\label{final_Fx_inter}
\end{eqnarray}
The integral in (\ref{final_Fx_inter}) may be expressed in terms of gamma 
functions: it equals $\alpha \beta \Gamma(\beta/\alpha)$. Therefore
\begin{equation}
{\cal F}_{x} = 2D\,(2\pi)^{1/2}\,
\left( \frac{2\gamma D}{\alpha^{2}} \right)^{\beta/2\alpha}\,
\frac{\Gamma \left( \frac{\alpha + \beta}{2\alpha} \right)}
{\left[ \Gamma(\beta/\alpha) \right]^{1/2}}\,\left[ 1 + O(D) \right]\,.
\label{final_Fx}
\end{equation}
The result (\ref{final_Fx}) only depends on the four parameters 
$\alpha, \beta, \gamma$ and $D$, as we would expect. An exactly analogous 
calculation of the total flux through the $y$-valley at $Y$ gives the same 
form but with $\alpha$ and $\beta$ interchanged:
\begin{equation}
{\cal F}_{y} = 2D\,(2\pi)^{1/2}\,
\left( \frac{2\gamma D}{\beta^{2}} \right)^{\alpha/2\beta}\,
\frac{\Gamma \left( \frac{\alpha + \beta}{2\beta} \right)}
{\left[ \Gamma(\alpha/\beta) \right]^{1/2}}\,\left[ 1 + O(D) \right]\,.
\label{final_Fy}
\end{equation}

Throughout the analysis we have ignored constant factors which multiply 
both the results coming from paths ending up in the $x$-valley and those 
ending up in the $y$-valley. The reason given was that eventually we would 
normalize the probabilities of state selection and therefore common factors
were irrelevant. We are now at the point where we can impose this
normalization, but before doing so, let us put in the constant multiplying
factor explicitly, and in doing so absorb the additional factors of
$2D\,(2\pi)^{1/2}$ which appear in (\ref{final_Fx}) and (\ref{final_Fy}) into
it. We therefore the final forms for the total fluxes along the $x$- and 
$y$-valleys as
\begin{eqnarray}
{\cal F}_{x} & = & {\cal K}\,
\left( \frac{2\gamma D}{\alpha^{2}} \right)^{\beta/2\alpha}\,
\frac{\Gamma \left( \frac{\alpha + \beta}{2\alpha} \right)}
{\left[ \Gamma(\beta/\alpha) \right]^{1/2}}\,\left[ 1 + O(D) \right]\,,
\label{final_x} \\
{\cal F}_{y} & = & {\cal K}\,
\left( \frac{2\gamma D}{\beta^{2}} \right)^{\alpha/2\beta}\,
\frac{\Gamma \left( \frac{\alpha + \beta}{2\beta} \right)}
{\left[ \Gamma(\alpha/\beta) \right]^{1/2}}\,\left[ 1 + O(D) \right]\,.
\label{final_y}
\end{eqnarray}
where ${\cal K}$ is the overall constant.

The relative probability of flowing into an $x$-valley has already been 
introduced in Sec. III. In terms of the total fluxes it is given by
\begin{equation}
N_{x} = \frac{{\cal F}_{x}}{{\cal F}_{x} + {\cal F}_{y}}\,,
\label{def_Nx}
\end{equation}
with $N_{y} = 1 - N_{x}$. From Eqns. (\ref{final_x}), (\ref{final_y}) and
(\ref{def_Nx}) we obtain:
\begin{equation}
N_{x} = \frac{1}{1 + \zeta(\alpha, \beta)\,\tilde{D}^{\sigma}}\,, \ \ \ 
N_{y} = \frac{1}{1 + \zeta(\beta, \alpha)\,\tilde{D}^{ - \sigma}}\,,
\label{Nx}
\end{equation}
where
\begin{eqnarray}
\tilde{D} & \equiv & \frac{2\gamma D}{\alpha \beta}\,, \ \ \ 
\sigma = \frac{\alpha^{2} - \beta^{2}}{2\alpha \beta}\,, \nonumber \\
\zeta(\alpha, \beta) & = & \left( \frac{\alpha}
{\beta} \right)^{(\alpha^{2} + \beta^{2})/2\alpha \beta}\,
\frac{\Gamma \left( \frac{\alpha + \beta}{2\beta} \right)}
{\Gamma \left( \frac{\alpha + \beta}{2\alpha} \right)}
\sqrt{\frac{\Gamma \left( \frac{\beta}{\alpha} \right)}
{\Gamma \left( \frac{\alpha}{\beta} \right)}}\,.
\label{Dzs}
\end{eqnarray}

Equations (\ref{Nx}) and (\ref{Dzs}) are the desired results. Note that the
only place where the coupling constant $\gamma$ enters is multiplying the
noise strength $D$. It is simple to see that this will always be the case
since, if $x$ and $y$ are measured in units of $X_{\rm min}$ and 
$Y_{\rm min}$ respectively in the reduced problem, the only place where 
$\gamma$ appears in the Langevin equation is multiplying $D$. Thus the
effect of the interaction, specified by $\gamma$, is to renormalize the
noise. This means that the probability of either the $x$-mode or of the 
$y$-mode being selected depends only on the three quantities $\alpha, \beta$
and $\gamma D$. It is clear that $N_x$ and $N_y$ given by (\ref{Nx}), have 
the right behavior in various limits. First of all, if $\alpha = \beta$ 
they are equal. If $\alpha$ and $\beta$ are not approximately the same, the
relative magnitudes of $N_{x}$ and $N_{y}$ are largely governed by their
dependence on $\tilde{D}$: if $\alpha > \beta$, then $\sigma > 0$, and
$N_{x} \alt 1$ while $N_{y} \alt \tilde{D}^{\sigma} \ll 1$. A more detailed 
comparison with the results of Monte Carlo simulations is given in Fig. 6.

\begin{figure}[tbh]
\begin{center}
\leavevmode
\hbox{\epsfxsize=0.8\columnwidth \epsfbox{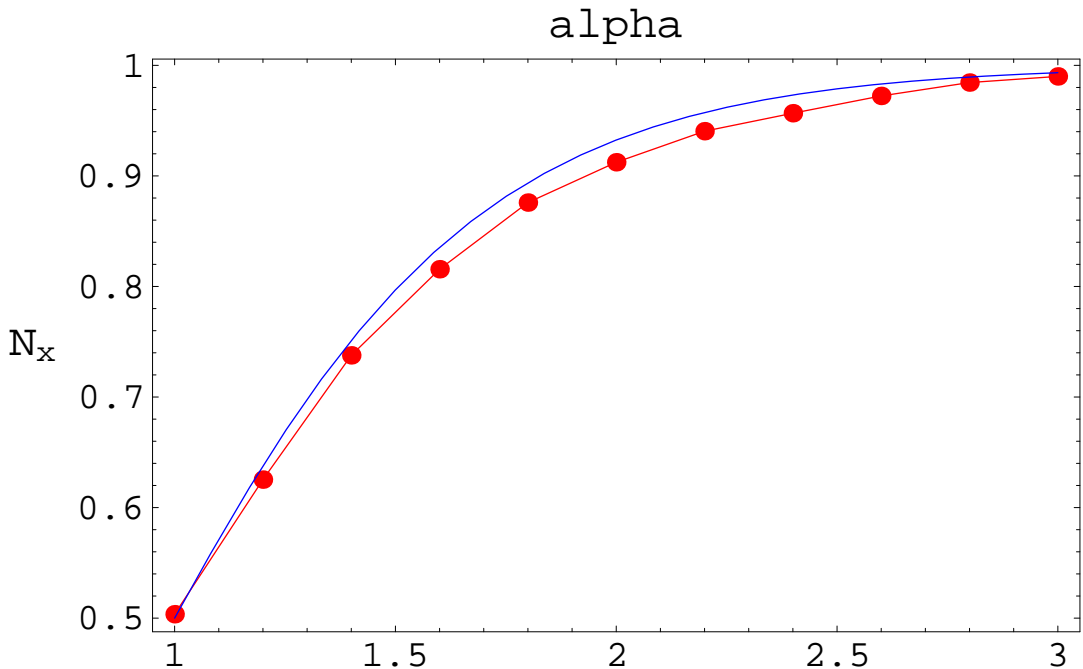}}
\end{center}
\caption{Probability of flowing into an $x$-valley as a function of $\alpha$,
with $\beta = 1$, $\gamma = 1$ and $D = 0.01$. The dots show simulation 
results and the full curve is our theoretical result for $N_{x}$ given by
(\ref{Nx}).}
\label{Fig6}
\end{figure}

The final result (\ref{Nx}) is in good, but not perfect, agreement with 
simulations. We believe that the only restriction on our method is that $D$ 
should be small, so that the method of steepest descent is appropriate. For
$D=0.01$, the neglect of $O(D)$ corrections to (\ref{final_x}) and
(\ref{final_y}) should mean that our result is within approximately 1\% of 
the simulation results. It is not clear from an examination of Fig. 6 
whether the slight discrepancy is due to the $O(D)$ corrections or to 
additional contributions which have not been accounted for. Since the 
agreement is still not perfect for smaller $D$, we should take the latter
explanation seriously and carry out a search to see if we can find such
additional terms. A possible source is the following. We have conjectured 
that the solution of (\ref{fullEL_x}) and (\ref{fullEL_y}) which has the 
least action can be obtained first by finding the $x_{c}(t)$ solution with 
$y_{c}=0$, and then linearizing about it. However, there are undoubtedly 
higher action solutions of (\ref{fullEL_x}) and (\ref{fullEL_y}) which have a 
non-trivial structure in $x$ and $y$, that is, solutions which cannot be 
obtained as a simple expansion about the $y=0$ solution. In other 
applications of the steepest descent method, we would neglect these 
solutions as they would give contributions which were exponentially smaller 
than the contributions from the least action solutions, and thus be 
completely negligible for small $D$. However, in this problem, after 
integrating over $T$, the previously exponentially small contribution 
becomes a power law (see Eqn. (\ref{time_int})), or more precisely, $D$ 
divided by the a quantity related to the action of the solution, all 
raised to a power. Although we do not know what power the higher action 
solutions will be raised to without further analysis, it is clear that 
solutions to (\ref{fullEL_x}) and (\ref{fullEL_y}) with, for instance, 
twice the action of the solution considered in this paper, have at least 
the potential of changing our result by a not insignificant amount. Thus 
it appears that the neglect of higher action solutions, which is common 
in most problems involving optimal paths, may not be so well-founded in 
the situation which we are considering in this paper. 

\section{conclusions}

In this paper we have presented the motivation, justification and 
calculational details of a method for determining the probabilities of 
various metastable states being selected for occupation, when an unstable 
state decays. Although, as we hope we have made clear, this is a widespread 
phenomenon, theoretical progress seems to have been hampered by a lack of 
sufficiently powerful tools with which to attack the problem. As a
consequence, most studies were carried out some time ago on 
one-dimensional systems (or quasi-one-dimensional systems). Yet, these are
actually the systems of least interest. The really interesting aspects of
the problem arise in two or more dimensions, where the system can 
perform Brownian motion near an unstable state, and then select a final
state through a mixture of randomness (noise) and dynamics (which will
necessarily be non-linear in all non-trivial cases). It is this interplay 
of randomness and determinism, both essential to the process, which makes 
the problem so difficult.

Our main purpose in this paper has been to show how the use of path-integral 
techniques, and especially the notion of optimal paths, can be used to
analyze this technically hard problem. In order to stress that relatively 
naive approaches are not able to capture much of the subtlety of this 
problem, we carried through what seemed to us the most simple-minded 
approach to the problem in two dimensions, and showed that it was not 
able to give satisfactory results. Although the extent of the agreement 
with numerical work is always open to interpretation, there were two 
features which signified that the naive method was defective. Firstly, 
there was no systematic way to proceed with 
the calculation; one was faced with {\it ad hoc} choices which had to be 
made at several junctures in the calculation. Secondly, the crude 
approximation (\ref{naive1}) to the already unsystematic result 
(\ref{naive2}), actually gives better agreement with simulations than 
(\ref{naive2}) itself, which indicates the dubious nature of these
results. It should also be borne in mind that the prediction that 
$N_{x} = 1/2$ when $\alpha = \beta$ is assured by symmetry and that in almost
all schemes $N_{x} \rightarrow 1$ as $\alpha$ gets very much larger than 
$\beta$, simply because the noise will have very little effect in this case.
If we add to this the expectation that $N_{x}(\alpha)$ will be a smooth 
function, we see that even the most simple minded approximation cannot be 
expected to be too far out. These observations indicate that we should 
focus more on the systematics (or lack thereof) of the calculation, rather 
than getting perfect agreement with simulation data.

The nature of the optimal paths used in the treatment presented here is
different from those in the more familiar problem of escape from one 
potential well to the other. In that situation it takes an exponentially 
large time ($\exp{\Delta V/D}$, where $\Delta V$ is the height of the 
barrier), to make the transition from one potential well to another, and this 
is reflected in the fact that the optimal paths are essentially of infinitely 
long duration ($T \rightarrow \infty$). In the problem studied in this paper
we are focusing on phenomena on much shorter time scales: the decay of a 
state with no barrier to a metastable state. Any subsequent process 
(presumably a noise-activated escape to a different well, of the kind we have
just discussed) is of no interest to us here. Actually, as explained in 
Sec. II, we do not even ask that the end point of the optimal path is a
metastable state, but only that it lies somewhere in the valley which leads 
to the metastable state in the full problem. The optimal paths in these 
cases are of finite duration. To achieve this, the initial velocity of the 
particle in the mechanical analogy has to be non-zero at the origin. As $T$ 
increases this initial velocity decreases, and eventually tends to zero if
$T$ tends to infinity.

Many of the other details of the calculation involving optimal paths turn out
to be remarkably subtle. For example, the use of the reduced potential 
simplifies the problem, but eliminating the actual minima means that we have
to specify an arbitrary final value of $x$ (if we are considering 
state-selection into the positive $x$-valley), which we denote by $X$. 
However, the probability of the positive $x$-valley being selected should not
depend on $X$, as long as $X$ has a value which lies in the valley. 
Reassuringly, this is what we find in our calculation, but it requires us
to take into account the $y$-dependence of the path for this
``common-sense'' condition to be observed. In our earlier treatment\cite{I},
we sought to give only the outline of our method, and made the assumption that
the total probability flux through a particular valley was proportional to (a)
its value on the axes, and (b) its maximum value, which occurs at 
$T = T^{*}$ --- the time for which $P(X, 0, 0|\vec{0}, 0)$ is a maximum at a
fixed $X$. Within this scheme there was no way of determining $X$ and $Y$,
and so we fixed them to our simulation results. The best values were 
$X \agt X_{\rm min}$ and $Y \agt Y_{\rm min}$, in line with our expectations. 
However, we have found here that it is possible to get a form for the 
$y$-component of the optimal path only if we assume that $X$ is 
sufficiently large --- which comes from the requirement that $T$ must be
sufficiently large --- and so setting $X \approx X_{\rm min}$ and
$Y \approx Y_{\rm min}$ is not possible in the approach we have adopted 
here. Needless to say, we expect that if we were able to find the full 
solution for the optimum path, then we would not need to constrain $X$ or 
$Y$ to be large: the expressions for the fluxes ${\cal F}_{x}$ and 
${\cal F}_{y}$ would be valid for all $X \ge X_{\rm min}$ and for all 
$Y \ge Y_{\rm min}$. Presumably the fact that ${\cal F}_{x}$ decreases from
its value at $X=X_{\rm min}$, and only becomes constant when $X$ equals a
few $X_{\rm min}$, is not due to a violation of flux conservation (this
would necessitate some flux leakage, and it is difficult to see where it 
would go to), but instead to the inapplicability of our calculational 
scheme for small $X$, which simply gives the incorrect expression for 
${\cal F}_{x}$. It should be stressed that this is not a problem for our
method: we simply take $X$ and $Y$ large enough so that ${\cal F}_{x}$ and
${\cal F}_{y}$ respectively, do tend to asymptotic values. These are the 
required values and are those which we use in our determination of $N_{x}$ 
and $N_{y}$.
     
An even more subtle aspect of the calculation concerns the applicability 
of the linear approximation in determining the $y$-component of the optimal 
path. Although at first sight $y_{c}(t)$ is not small for large $X$, it turns 
out that the range of final values of $y_{c}(t)$, denoted by $y_{f}$,
decreases so fast as $X$ increases, that the small $y_{c}(t)$ approximation 
is still valid. Put more intuitively: if $T$ is large, the path has to take a 
wide detour to large $y$ if it is to take sufficiently large time to reach
$(X, y_{f})$ (recall that $T$ is fixed for these paths). But ``large $y$''
means large relative to $y_{c}(T)=y_{f}$, and, as discussed earlier, the
current ${\cal J}_{x}(X, y_{f}, T)$ is effectively non-zero only for tiny 
values of $|y_{f}|$. So while it is true that on the scale of $y_{f}$ the 
excursions of the optimal path spread deep into the plane, in fact the 
range of $y_{f}$ is so minute, that the linearization assumption remains 
valid. The result that the probability is so concentrated about the axes is 
used in other places in the calculation. For example, in Sec. VI we argued 
that the ``volume'' integral in (\ref{intermed_step1}) could be replaced by 
the ``surface'' integral in (\ref{intermed_step2}), and then we proceeded 
to ignore some of the contributions from the ``surfaces'' in the later 
integral. We specified the volume in the case of the positive $x$-valley as 
the entire plane to the right of the line $x=X$. Actually, this is not 
quite correct, since we need to define a similar volume for the $y$-valleys, 
and these will overlap in the sectors where $x$ and $y$ are both large. We 
can be reasonably vague about how precisely to define these volumes and 
surfaces simply because ${\cal J}_{x}$ (and ${\cal J}_{y}$) fall away so 
fast that they have an utterly negligible contribution however we define 
them. In any reasonable definition, the only contribution to the surface 
integral (\ref{intermed_step2}) will be on the line $x=X$, very close to 
the $x$-axis.

We believe that one of the best indicators of the correctness of our approach
is the cancellation of the various factors of $X$ coming from several 
different sources: (a) the fluctuations about the $y_{c}(t)$-solution, (b) 
the action of the $y_{c}(t)$-solution (after integration over $y_{f}$), (c) 
the action of the $x_{c}(t)$-solution (after integration over $T$), and
(d) the definition of the current in terms of the probability distribution.
What remains after these cancellations is essentially independent of $X$,
as would intuitively be expected. As we have repeatedly stressed, many of
these contributions come from assuming that $T$ is large. This seems somewhat
paradoxical, since we would expect that state selection is determined at 
earlier times. This can be understood if one realizes that the large $X$ 
calculation determines the normalization of the flux in (\ref{final_FxX}), or
equivalently, the quantity $\zeta(\alpha, \beta)$ in (\ref{Nx}). To obtain
a finite $X \rightarrow \infty$ limit, it is essential to include a 
non-trivial $y_{c}(t)$ optimal path in the analysis. If the optimal path
is taken to be $y_{c}(t)=0$, then only the $\tilde{D}^{\sigma}$ factor in
(\ref{Nx}) is determined. This was essentially the approach we adopted in 
our earlier paper\cite{I}. While we believe that the treatment given here 
is a great improvement on that reported in Ref. \cite{I}, there is no doubt
room for further improvement. For instance, the linearization approximation 
seems very reasonable on physical grounds, but it would be useful to put
it on a sounder mathematical footing. A deeper understanding of the origin 
of the small scale set by $\exp{(-\gamma X^{2}/\alpha)}$ would also be
valuable.

The competition between new modes, when a previously stable mode becomes
unstable, is responsible for much of the emergent order found in systems 
far from equilibrium. The continual branching to more complex structures 
that this entails\cite{arecchi}, is not only governed by the dynamics of 
the process but also by the random fluctuations, or noise, generated by 
the large number of other degrees of freedom of the system not explicitly 
included in the description. The model which we have investigated in this 
paper is, as has already been emphasized, the simplest showing enough of 
the complicated features of this process of state selection to serve as an 
illustrative example of our method. In forthcoming papers we will show how 
the method can be applied to more complex examples, such as state selection 
in lasers and the study of population dynamics in a fluctuating environment. 
We believe that the ideas and techniques developed here will not only be 
applicable to these situations, but to many others where multiple states 
compete for occupation. 

\acknowledgments 
AM wishes to thank the Department of Physics at the University of Illinois at
Urbana-Champaign, where part of this work was carried out, for hospitality, 
and EPSRC for financial support under Grant No. K/79307. MT gratefully 
acknowledges support from the MRSEC program of the NSF under award number
DMR 9808595.
 
\appendix 
\section{} 
 
In this appendix, we explore the consequences of the elementary theory 
put forward at the end of Sec. III. We will obtain simple expressions 
for the probability that the particle ends up in an $x$-valley, which are 
compared to Monte Carlo simulations in Sec. III.

We begin with the first ingredient in this simple theory. Since $\gamma = 0$ 
we have only to solve the Langevin equations 
$\dot{x} = \alpha x + \eta_{x}(t)$ and $\dot{y} = \beta y + \eta_{y}(t)$. 
Such linear problems can always be exactly solved. Since $x$ and $y$ are 
linearly related to $\eta_{x}$ and $\eta_{y}$, they are also Gaussian random 
variables. It is easy to show that 
$\langle x(t) \rangle = \langle y(t) \rangle = 0$ and therefore the
probability that the particle is at $(x,y)$ at time $t$, given that it
started at the origin at time $t=0$ is
\begin{equation}
P(x, y, t|0, 0, 0) = {\cal N}\exp\left\{ - 
\frac{x^{2}}{2\langle x^{2}(t) \rangle} -
\frac{y^{2}}{2\langle y^{2}(t) \rangle} \right\}\,,
\label{naive_prob}
\end{equation}
where ${\cal N}$ is a normalization constant. For later use we will also 
need 
\begin{equation}
\langle x^{2}(t) \rangle = \frac{D}{\alpha}
\left( e^{2\alpha t}-1\right)\ \ ; \ \ 
\langle y^{2}(t) \rangle = \frac{D}{\beta}\left( e^{2\beta t}-1\right)\,,
\label{naive_variances}
\end{equation}

To quantify the second ingredient, first suppose that the coordinates of the
particle $(x, y)$ are in the first quadrant. If the angle it makes with the
positive $x$-axis, $\tan^{-1}(y/x)$ is less than 
$\tan^{-1}(Y_{\rm min}/X_{\rm min})$, then according to our criterion, it 
goes into the $x$-valley. Since the tan function is monotonic in the
interval $(0, \pi/2)$, and generalizing to other quadrants by using $|x|$
and $|y|$ rather than $x$ and $y$, the criterion becomes:
\begin{eqnarray}
\frac{|x|}{X_{\rm min}} & > & \frac{|y|}{Y_{\rm min}}\ \ \Rightarrow \ \
{\rm particle \ chooses \ } x{\rm -valley}
\label{naive_xchoice} \\ \nonumber \\
\frac{|y|}{Y_{\rm min}} & > & \frac{|x|}{X_{\rm min}}\ \ \Rightarrow \ \
{\rm particle \ chooses \ } y{\rm -valley}\,. 
\label{naive_ychoice}
\end{eqnarray}

These ingredients can now be put together. The question reduces to how 
often the particle is in the sector specified by (\ref{naive_xchoice}) at a
given time $t$ and how often it is in the sector specified by 
(\ref{naive_ychoice}) at this time. This probability is found by integrating 
(\ref{naive_prob}) over all $x$ and all $y$ with the constraint that it 
satisfies either (\ref{naive_xchoice}) or (\ref{naive_ychoice}) depending on
whether we want the probability of ending up in the $x$-valley or $y$-valley.
Restricting ourselves to the positive quadrant, which we can obviously do on
symmetry grounds, the probability of ending up in an $x$-valley, $N_{x}$
and in a $y$-valley, $N_{y}$, are
\begin{eqnarray}
N_{x} & = & 4{\cal N}\int^{\infty}_{0}dx\,\int^{\lambda x}_{0}dy
\exp\left\{ - \frac{x^{2}}{2\langle x^{2}(t) \rangle} -
\frac{y^{2}}{2\langle y^{2}(t) \rangle} \right\}\ \ {\rm and} 
\label{naive_Nx1} \\ \nonumber \\
N_{y} & = & 4{\cal N}\int^{\infty}_{0}dy\,\int^{\lambda^{-1} y}_{0}dx
\exp\left\{ - \frac{x^{2}}{2\langle x^{2}(t) \rangle} -
\frac{y^{2}}{2\langle y^{2}(t) \rangle} \right\}\,, 
\label{naive_Ny1} 
\end{eqnarray}
where the factor $4$ comes from including the contributions from the other
quadrants and $\lambda = Y_{\rm min}/X_{\rm min}$. In the $N_{x}$ integral, 
change variables from $y$ to $z$ where $y=\lambda xz$ at fixed $x$. The $x$ 
integral can now be easily performed, giving
\begin{equation}
N_{x} = 4{\cal N}\frac{Y_{\rm min}}{X_{\rm min}} \langle x^{2}(t) \rangle
\int^{1}_{0} \frac{dz}{1 + \kappa^{2}(t)z^{2}}\,,
\label{naive_Nx2}
\end{equation}
where
\begin{equation}
\kappa^{2}(t) = \frac{\lambda^{2} \langle x^{2}(t) \rangle}
{\langle y^{2}(t) \rangle} =  \frac{Y^{2}_{\rm min} \langle x^{2}(t) \rangle}
{X^{2}_{\rm min}\langle y^{2}(t) \rangle}\,.
\label{kappa_def}
\end{equation}
In a similar way
\begin{equation}
N_{y} = 4{\cal N}\frac{X_{\rm min}}{Y_{\rm min}} \langle y^{2}(t) \rangle
\int^{1}_{0} \frac{dz}{1 + \kappa^{-2}(t)z^{2}}\,.
\label{naive_Ny2}
\end{equation}
From (\ref{naive_variances}) we see that if $\alpha$ and $\beta$ are not 
very different from each other, then $\kappa(t)$ will be neither very large 
nor very small, and we can ignore the integrals in (\ref{naive_Nx2}) and 
(\ref{naive_Ny2}) as a first approximation. Then determining ${\cal N}$ by
asking that $N_{x} + N_{y} = 1$ and ignoring the factor $1$ compared to
the exponentials in (\ref{naive_variances}) yields
\begin{equation}
N_{x} = \frac{e^{2\alpha t}}{e^{2\alpha t} + e^{2\beta t}}\ \ ; \ \
N_{y} = \frac{e^{2\beta t}}{e^{2\alpha t} + e^{2\beta t}}\,.
\label{naive_form1}
\end{equation}
There are several simple arguments which give formulas resembling this. It
is not difficult to improve on this by including the integrals in 
(\ref{naive_Nx2}) and (\ref{naive_Ny2}). Once again determining ${\cal N}$ by
normalization gives
\begin{equation}
N_{x} = \frac{2}{\pi}\tan^{-1} [ \kappa (t) ]\ \ ; \ \
N_{y} = \frac{2}{\pi}\tan^{-1} [ \kappa^{-1}(t) ]\,,
\label{naive_form2}
\end{equation}
where we have used the result $\tan^{-1}(w) + \tan^{-1}(1/w) = \pi/2$. If
$\langle x^{2}(t) \rangle \gg \langle y^{2}(t) \rangle$ or
$\langle y^{2}(t) \rangle \gg \langle x^{2}(t) \rangle$, which can happen 
even when $\alpha$ and $\beta$ are not too different, we may replace
(\ref{naive_form2}) by a simpler form which resembles (\ref{naive_form1}),
but with constants multiplying the exponentials. 

The simple result (\ref{naive_form1}) does not depend on the parameter 
$\gamma$, which represents the non-linear part of the potential, or on
the noise strength, $D$. This is also true of (\ref{naive_form2}) if we
ignore the factor $1$ compared to the exponentials in (\ref{naive_variances}).
To understand this, let us suppose that $\alpha > \beta$. From 
(\ref{naive_form1}) or (\ref{naive_form2}), we see that when $t$ is small
although there is a slight bias in favor of the $x$-valley over the 
$y$-valley, it is not that marked. However, as $t$ increases the asymmetry
becomes more marked, until at large times the particle has only a very
small chance of being in the $y$-valley. We need to simulate the role that
the saddle points play in state selection by choosing the time to have the
value $t_{c}$, which is the time at which it is most likely that the $x$ and 
$y$ coordinates of the particle will have magnitudes $X_{\rm min}$ and
$Y_{\rm min}$ respectively. This is implemented by asking that 
$\langle x^{2}(t_{c}) \rangle = X^{2}_{\rm min}$ or
$\langle y^{2}(t_{c}) \rangle = Y^{2}_{\rm min}$. Note that we have used the
word ``or'', since only one relation is required to determine $t_{c}$, and 
the two relations given give incompatible values for $t_{c}$ unless 
$\alpha = \beta$. Once again, as is common in many aspects of the naive 
approach to state selection which we have been summarizing in this appendix, 
there is a considerable degree of arbitrariness in the choice of $t_{c}$.
We will assume that for $\alpha > \beta$, when the particle is most likely
to go into the $x$-valley, $t_{c}$ is determined from the condition on 
$\langle x^{2}(t_{c})$ and vice-versa. Then from (\ref{naive_variances}),
\begin{equation}
t_{c} = - \frac{1}{2\alpha}\ln \left( \frac{\gamma D}{\alpha\beta} \right)\ \
; \ \ \alpha \geq \beta ,
\label{naive_tc}
\end{equation}
which has the characteristic form $-\ln D$ as $D \rightarrow 0$. 

If we substitute (\ref{naive_tc}) into (\ref{naive_form1}) we obtain the
probability of ending up in an $x$-valley. In terms of the quantities
\begin{equation}
\hat{D} \equiv \frac{\gamma D}{\alpha\beta} \ \ {\rm and} \ \
\rho = \frac{(\alpha - \beta)}{\alpha}\,,
\label{naive_Dandrho}
\end{equation}
we obtain
\begin{equation}
N_{x} = \frac{1}{1 + \hat{D}^{\rho}}\,.
\label{naive_form3}
\end{equation}
If we substitute (\ref{naive_tc}) into (\ref{naive_form2}) we obtain 
the more general form
\begin{equation}
N_{x} = \frac{2}{\pi} \tan^{-1} \hat{D}^{-\rho}\,.
\label{naive_form4}
\end{equation} 

Equations (\ref{naive_form3}) and (\ref{naive_form4}) are the principal 
results of this appendix. The two main points we wish to make is (i) their
derivation is somewhat arbitrary --- there are several other similar
assumptions we could make which would give us slightly different formulas,
and (ii) although, as is discussed in Sec. III, they show the right 
qualitative features, they are are not in good agreement with Monte Carlo 
simulations. Thus a more systematic and sophisticated theory is required. 
This is given in Secs. IV, V and VI of the paper.

\section{}

The purpose of this appendix is to explore the solution of (\ref{EL_y}), where
$x(t)$ is given by (\ref{x_soln}) and where $y(0) = 0$ and $y(T) = y_{f}$.
Throughout we will assume that $T$ is large in the sense that
$e^{-\alpha T} \ll 1$ and $e^{-\beta T} \ll 1$, which will allow us to 
obtain a rather simple form for the solution. 

\underline{Small $t$ solution.} Let $\zeta = X^{2}{\rm cosech}^{2}\alpha T$.
Since $e^{- \alpha T} \ll 1$, $\zeta \ll 1$. Therefore (\ref{EL_y}) becomes
\begin{equation}
\ddot{y} = y \{ \beta^{2} - 2\zeta \gamma (\alpha + \beta )\sinh^{2}\alpha t
+ O(\zeta^{2}) \}\,.
\label{zeta_expansion}
\end{equation}
We look for a solution of (\ref{zeta_expansion}) as a power-series in
$\zeta$:
\begin{equation}
y(t) = y_{0}(t) + y_{1}(t) \zeta + O(\zeta^{2})\,.
\label{power_series}
\end{equation}
Clearly the zeroth-order solution which satisfies $y(0) = 0$ is 
$y_{0}(t) = A\sinh \beta t$, where $A$ is an arbitrary constant. It is
straightforward to determine $y_{1}(t)$, but since a short analysis shows that
$y_{1}(t)\zeta \ll y_{0}(t)$ for all $t \alt T/2$, we conclude that, it is 
sufficient to take the small $t$ of the solution of (\ref{EL_y}) --- which
we denote by $y_{<}$ --- to be 
\begin{equation}
y_{<}(t) = A\sinh \beta t \ \ ; \ \ t < t_{m}
\label{smallt}
\end{equation}

\underline{Large $t$ solution.} In this regime both $e^{-\alpha T} \ll 1$ and 
$e^{-\alpha t} \ll 1$, and therefore $x_{c}^{2}(t)$ and $x_{c}^{4}(t)$ may be
approximated by $X^{2}e^{-2\alpha (T - t)}$ and $X^{4}e^{-4\alpha (T - t)}$
respectively. Using these forms in (\ref{EL_y}) and making the change of
variables 
\begin{equation}
z = \frac{\gamma X^{2}}{\alpha}\,e^{-2\alpha (T - t)}\ \ ; \ \ 
f = e^{z/2}\,z^{-\beta/2\alpha} y\,,
\label{CofV}
\end{equation}
we find that $f(z)$ satisfies the equation
\begin{equation}
z \frac{d^{2}f}{dz^{2}} + (\nu - z) \frac{df}{dz} = 0\ \ ; \ \ 
\nu = \frac{(\alpha + \beta)}{\alpha}\,.
\label{f_eqn}
\end{equation}
This equation is easily solved to give
\begin{equation}
f(z) = {\cal A} + {\cal B} \int^{\gamma X^{2}/\alpha}_{z} dz\, z^{-\nu}\, 
e^{z}\,,
\label{full_fsoln}
\end{equation}
where ${\cal A}$ and ${\cal B}$ are arbitrary constants and where the upper
limit of the $z$-integral has been arbitrarily chosen to be the value that
$z$ takes on when $t=T$. 

We now have to match (\ref{full_fsoln}) to (\ref{smallt}) at $t = t_{m}$. It
will turn out that we do not have to specify $t_{m}$ precisely, but let us
assume for now that it has the value $T/2$. Then the corresponding value of
$z$, $z_{m}$ is seen from (\ref{CofV}) to be very small. Formally the
matching procedure consists of equating both $y(t_{m})$ and $\dot{y}(t_m)$ 
for the solutions in the large and small time regimes. We will carry out this
procedure below. However, let us first give a quick argument which gives
us the correct final result.  

For $t$ near $t_{m}$, $y_{<}(t) \sim A\,e^{\beta t}/2$. Also from 
(\ref{CofV}), $y_{>}(t) = e^{\beta t}\,e^{-z/2}\,f(z)$, and since 
$z_{m} \ll 1$, it follows that for $t$ near $t_{m}$, 
$y_{>}(t) \sim e^{\beta t}\,f(z)$. To get a smooth match, we require $f(z)$
to be a constant, $A/2$, for values of $z$ near $z_{m}$. Looking now at
(\ref{full_fsoln}), we see that the first term is constant, but the second
term changes very quickly in the small $z$ regime, since it diverges in the
$z \rightarrow 0$ limit. We therefore take ${\cal B} = 0$ to get a smooth 
match, which in turn gives ${\cal A} = A/2$. Therefore
\begin{equation}
y_{>}(t) = \frac{A}{2}\,e^{\beta t}\,
\exp\left\{-\frac{\gamma X^{2}}{2\alpha}\,e^{-2\alpha(T - t)} \right\}\ \ ;
\ \ t > t_{m}\,.
\label{larget}
\end{equation}
Having matched (\ref{smallt}) and (\ref{larget}) near $t_{m}$, we can now
fix the constant $A$ from the boundary condition $y(T) = y_{f}$. Doing
this gives us the form (\ref{y_soln}). 

Let us briefly indicate how we arrive at this result in a more systematic
fashion. We can choose to either match $\{ f(z), f'(z) \}$ at $z=z_{m}$ or 
$\{ y(t), \dot{y}(t) \}$ at $t=t_{m}$. They are related by
\begin{eqnarray}
e^{-\beta t}\,y(t) & = & e^{-z/2}\,f(z)\ \ {\rm and} 
\label{yandf} \\ 
\frac{\dot{y}(t)}{y(t)} = \beta & - & \alpha z + 2\alpha z 
\frac{f'(z)}{f(z)}\,. 
\label{ydotandfprime}
\end{eqnarray}
Now, by successive integration by parts, we can obtain an approximation to
the integral in (\ref{full_fsoln}) valid for small $z$. For our purposes
we need retain only the first term:
\begin{equation}
f(z) \approx {\cal A} + {\cal B}\,\frac{z^{-\nu + 1}\,e^{z}}{\nu - 1}\ \ ;
\ \ z \approx z_{m}\,.
\label{approx_fsoln}
\end{equation}
Therefore to leading order
\begin{eqnarray}
f_{>}(z_{m}) & = &  {\cal A} + {\cal B}\,
\frac{\alpha z_{m}^{-\beta/\alpha}\,e^{z_{m}}}{\beta}\ \ {\rm and}
\nonumber \\
f_{>}'(z_{m}) & = &  - {\cal B}\,e^{z_{m}}\,z_{m}^{-(\alpha + 
\beta)/\alpha}\,.
\label{fgreater}
\end{eqnarray}
On the other hand $y_{<}(t_{m}) = (A/2) e^{\beta t_{m}}$ and 
$\dot{y}_{<}(t_{m}) = (\beta A/2) e^{\beta t_{m}}$. So from (\ref{yandf})
and (\ref{ydotandfprime})
\begin{equation}
f_{<}(z_{m}) = \frac{A}{2} e^{z_{m}/2} \ \ ; \ \ 
f_{<}'(z_{m}) = \frac{A}{4} e^{z_{m}/2}\,.
\label{flessthan}
\end{equation}
Matching (\ref{fgreater}) and (\ref{flessthan}) gives
\begin{equation}
{\cal A} = \frac{A}{2}\left[ 1 + O(z_{m}) \right]\ \ ;\ \ 
{\cal B} = \frac{A}{4}\,z_{m}^{(\alpha + \beta)/\alpha}\,
\left[ 1 + O(z_{m}) \right]\,.
\label{matching}
\end{equation}
Substituting (\ref{matching}) into (\ref{full_fsoln}) we obtain
\begin{equation}
f_{>}(z) = \frac{A}{2}\,\left\{ 1 - z_{m}^{(\alpha + \beta)/\alpha}\,
\int^{\gamma X^{2}/\alpha}_{z} dz\,z^{(\alpha + \beta)/\beta}\,
e^{z} \right\}\,.
\label{explicit_f}
\end{equation}
Clearly for values of $z$ that are not small, the second term is completely 
negligible and so $f_{>}(z) \sim A/2$ to a very good approximation. However, 
even for values of $z$ near to $z_{m}$ we can neglect this term --- it 
gives a contribution which is $O(z_{m})$. Therefore, once again, we 
obtain (\ref{larget}).

In determining the form of $y(t)$, we have implicitly assumed that the 
solution of least action is such that $y(t) \neq 0$ for $t \neq 0$. However, 
there are other solutions of the type $y(t) = 0$, for $t \leq t_{1}$, (where
$t_{1} < T$) and a non-trivial solution of (\ref{EL_y}), where $x(t)$ is 
given by (\ref{x_soln}) and where $y(t_{1}) = 0$ and $y(T) = y_{f}$, for 
$t \geq t_{1}$. Intuitively, we would expect that the solution with 
$t_{1} =0$, which is the one found above, is the solution of least action, 
and all those with $t_{1} > 0$ have greater action. This is due to the fact 
that, especially for large $t_{1}$, the path will have to sharply curve 
away from the $x$-axis if it is to satisfy the condition at $t=T$. We have 
checked this by numerically solving the differential equation (\ref{EL_y}) 
for $y(t)$ with different $t_{1}$ values, and found that the action 
monotonically increases with $t_{1}$.  

Let us end by outlining the evaluation of the classical action
(\ref{class_action}). To evaluate $E$ we choose to take $t=0$, since in
this case $U=0$. Also $V|_{t=0} = 0$ and 
$V|_{t=T} = - \frac{1}{2}\alpha X^{2} + \frac{1}{2} y_{f}^{2}[\gamma X^{2}
-\beta]$. Therefore
\begin{eqnarray}
S_{c} & = & \left\{ \frac{1}{2}\int^{T}_{0} dt\,\dot{x}^{2}_{c}
- \frac{1}{4} \left. \dot{x}^{2}_{c} \right|_{t=0}\,T
- \frac{1}{4} \alpha X^{2} \right\} \nonumber \\
& + & \left\{ \frac{1}{2}\int^{T}_{0} dt\,\dot{y}^{2}_{c} 
- \frac{1}{4} \left. \dot{y}^{2}_{c} \right|_{t=0}\,T + \frac{1}{4} 
y_{f}^{2}[\gamma X^{2} - \beta] \right\}\,.
\label{clas_action}
\end{eqnarray}
The first bracket on the right-hand side of (\ref{clas_action}) is easily
evaluated from $x_{c}(t)$ given by (\ref{x_soln}). One finds 
$(\alpha X^{2}/4)({\rm coth} \alpha T\, - 1)$. The second bracket
requires a little more calculation, since $y_{c}(t)$ has a different 
functional form depending on whether $t$ is less than, or greater than,
$t_{m}$. Substituting the form for $t \le t_{m}$ first of all, gives for
the second bracket:
\begin{equation}
\frac{1}{2} \int^{T}_{t_{m}}  dt\,\dot{y}^{2}_{c} + \frac{A^{2}\beta}{8}
\sinh 2\beta t_{m} - \frac{A^{2}\beta^{2}}{4} \left( T - t_{m} \right) 
+ \frac{1}{4} y_{f}^{2}[\gamma X^{2} - \beta]\,.
\label{first_expression}
\end{equation}
Using (\ref{y_soln}) and changing variables from $t$ to $z$ (given by
(\ref{CofV})) one finds
\begin{equation}
\frac{1}{2} \int^{T}_{t_{m}} dt\,\dot{y}^{2}_{c}(t) = \frac{A^{2}}{16\alpha}\,
\left( \frac{\alpha}{\gamma X^{2}} \right)^{\beta/\alpha} e^{2\beta T}\,
\int^{\gamma X^{2}/\alpha}_{z_{m}} dz\,z^{(\beta/\alpha) - 1}\, e^{-z}\,
(\beta - \alpha z)^{2}\,.
\label{second_expression}
\end{equation}
The range of the integral in (\ref{second_expression}) may be taken as
$(0, \gamma X^{2}/\alpha)$ as long as we subtract out 
\begin{equation}
\int^{z_{m}}_{0} dz\,z^{(\beta/\alpha) - 1}\, e^{-z}\,
(\beta - \alpha z)^{2} = \alpha \beta z_{m}^{\beta/\alpha}\,\left[ 1 +
O(z_{m}) \right]\,.
\label{third_expression}
\end{equation}
The upper limit in the integral in (\ref{second_expression}) is $O(1)$,
and therefore the integral itself is also $O(1)$. Therefore the contribution
(\ref{third_expression}) is negligible compared to it, and so we may 
effectively replace the lower limit of the integral in 
(\ref{second_expression}) by zero. The third term in (\ref{first_expression})
being linear in $(T - t_{m})$ is negligible compared with the second term
which is exponential in $2\beta t_{m}$. This term is in turn negligible 
compared to (\ref{second_expression}), since $e^{-2\beta (T - t_{m})} \ll 1$.
So the leading contribution to (\ref{second_expression}) is 
\begin{equation}
\frac{1}{4} y_{f}^{2}[\gamma X^{2} - \beta] + \frac{A^{2}}{16\alpha}\,
\left( \frac{\alpha}{\gamma X^{2}} \right)^{\beta/\alpha} e^{2\beta T}\,
\int^{\gamma X^{2}/\alpha}_{0} dz\,z^{(\beta/\alpha) - 1}\, e^{-z}\,
(\beta - \alpha z)^{2}\,.
\label{fourth_expression}
\end{equation}
Substituting the value of $A$ determined from the boundary condition
$y(T)=y_{f}$, into (\ref{fourth_expression}), the classical action 
(\ref{clas_action}) becomes
\begin{eqnarray}
S_{c} & = & \frac{\alpha X^{2}}{4}\,({\rm coth} \alpha T\, - 1) + 
\frac{1}{4} y_{f}^{2}[\gamma X^{2} - \beta] \nonumber \\
& + & \frac{y_{f}^{2}}{4 \alpha}\,\left( \frac{\alpha}
{\gamma X^{2}} \right)^{\beta/\alpha}\,\exp\left\{ \frac{\gamma X^{2}}
{\alpha} \right\}\,\int^{\gamma X^{2}/\alpha}_{0} dz\,
z^{(\beta/\alpha) - 1}\, e^{-z}\,(\beta - \alpha z)^{2}\,.
\label{cla_action}
\end{eqnarray}

\end{document}